\documentclass{article}
\usepackage{amssymb}
\usepackage{fleqn}
\usepackage{epsfig}
\usepackage{graphicx}
\usepackage{amsmath}

\def\beq{\begin{equation}}
\def\eeq{\end{equation}}
\def\bea{\begin{eqnarray}}
\def\eea{\end{eqnarray}}
\def\bq{\begin{quote}}
\def\eq{\end{quote}}

\def\gappeq{\mathrel{\rlap {\raise.5ex\hbox{$>$}}
{\lower.5ex\hbox{$\sim$}}}}
\def\lappeq{\mathrel{\rlap{\raise.5ex\hbox{$<$}}
{\lower.5ex\hbox{$\sim$}}}}
\def\Toprel#1\over#2{\mathrel{\mathop{#2}\limits^{#1}}}

\def\beq{\begin{equation}}
\def\eeq{\end{equation}}
\def\bea{\begin{eqnarray}}
\def\eea{\end{eqnarray}}
\def\bq{\begin{quote}}
\def\eq{\end{quote}}
\def\gappeq{\mathrel{\rlap {\raise.5ex\hbox{$>$}}
{\lower.5ex\hbox{$\sim$}}}}
\def\lappeq{\mathrel{\rlap{\raise.5ex\hbox{$<$}}
{\lower.5ex\hbox{$\sim$}}}}
\def\Toprel#1\over#2{\mathrel{\mathop{#2}\limits^{#1}}}

\newcommand{\bc}{\begin{center}}
\newcommand{\ec}{\end{center}}

\newcommand{\ba}{\begin{array}}
\newcommand{\ea}{\end{array}}

\begin{document}

\pagestyle{empty}
\begin{flushright}
ROME1/1434/06~

DSFNA/26/2006~~
\end{flushright}
\vspace*{15mm}

\begin{center}
\textbf{Semi-Inclusive $B$ Decays and a Model for Soft-Gluon Effects} \\[0pt]

\vspace*{1cm}

\textbf{Ugo Aglietti}\footnote{e-mail address: Ugo.Aglietti@roma1.infn.it} \\[0pt]

\vspace{0.3cm}
Dipartimento di Fisica,\\
Universit\`a di Roma ``La Sapienza'', \\
and I.N.F.N.,
Sezione di Roma, Italy. \\[1pt]

\vspace{0.3cm}
\textbf{Giancarlo Ferrera}\footnote{e-mail address: Giancarlo.Ferrera@roma1.infn.it} \\[0pt]

\vspace{0.3cm}
Dipartimento di Fisica,\\
Universit\`a di Roma ``La Sapienza'', \\
and I.N.F.N.,
Sezione di Roma, Italy. \\[1pt]

\vspace{0.3cm}
\textbf{Giulia Ricciardi}\footnote{e-mail address: Giulia.Ricciardi@na.infn.it} \\ [0pt]

\vspace{0.3cm}
Dipartimento di Scienze Fisiche,\\
Universit\`a di Napoli ``Federico II'' \\
and I.N.F.N.,
Sezione di Napoli, Italy. \\ [1pt]

$~~~$ \\[0pt]
\vspace*{2cm} \textbf{Abstract} \\[0pt]
\end{center}

We compare experimental spectra of radiative and semileptonic $B$
decays with the predictions of a model based on soft-gluon
resummation to next-to-next-to-leading order and on a ghost-less
time-like coupling. We find a good agreement with photon spectra
in the radiative decay and with hadron mass distributions in the
semileptonic one: the extracted values for
$\alpha_S\left(m_Z\right)$ are in agreement with the current PDG
average within at most two standard deviations. The agreement is
instead less good for the electron spectra measured by BaBar and
Belle in semileptonic decays for small electron energies ($ \le
2.2 $ GeV): our spectrum is harder. We also show that, in general,
the inclusion of next-to-next-to-leading order effects is crucial
for bringing the model closer to the data and that the non-power
expansion introduced in the framework of analytic coupling studies
does not accurately describe soft-gluon effects.

\vspace*{1cm} \noindent 

\vfill\eject

\setcounter{page}{1} \pagestyle{plain}

\section{Introduction}

The aim of this work is to analyze measured $B$ decay spectra
with a model based on $(i)$ soft gluon resummation to
next-to-next-to-leading order and $(ii)$ an effective QCD coupling having no
Landau pole \cite{shirkov}.
This coupling is constructed by means of an extrapolation at low energy
of the high-energy behavior of the standard coupling.
More technically, an analyticity principle is used.

$B$ decay spectra are substantially affected by long-distance effects,
the most important ones being the soft interactions occurring in the
fragmentation of the $B$ meson into the $b$ quark.
The $B$ meson --- a colorless composite particle ---
emits the spectator quark and radiates soft gluons,
\beq
B \, \to \, b \, + \, sp \, + \, g_1 \, + \, g_2 \, + \cdots \, + \, g_n,
\eeq
to convert into the colored $b$ quark which later decays
because of weak interactions,
\beq
b \, \to \, s \, + \, \gamma
\eeq
or
\beq
b \, \to \, u \, + \, l \, + \, \nu.
\eeq
Perturbation theory can describe the fragmentation of a $b$ quark
into a $b$ quark with a fraction of the original energy--momentum
as an effect of multiple gluon radiation, but it clearly cannot describe that
part of the fragmentation involving the spectator quark.
Physical intuition suggests that initial bound-state
effects are substantial for
\beq
\label{lasolita}
m_X^2 \, \approx \, m_B \, \Lambda_{QCD} \, \approx \, 2 \, {\rm GeV}^2,
\eeq
which is experimentally interesting: that is the well known
Fermi motion of the $b$ quark in the $B$ meson
($m_X$ is the final invariant hadron mass).
This non-perturbative effect --- which classically can be pictured
as a small vibration of the $b$ quark in the $B$ meson
because of the interactions with the spectator --- has been
formalized in an effective field theory by means of the well-known
shape function or structure function of the heavy flavors
\cite{Bigi:1993ex}.
Many models have been constructed to describe Fermi-motion as a
genuinely non-perturbative effect involving the hadron
structure \cite{Buchmuller:2005zv};
perturbative corrections are included, if desired, later on and play
in any case a minor role.
In this work we adopt a different philosophy:
in essence, we assume that the fragmentation of the lowest-lying beauty meson
into the beauty quark and the spectator quark can be described as a radiation
process off the $b$ with a proper coupling.
Even though dynamics of light degrees of
freedom in the $B$ meson is complicated, we assume that
the related effects on semi-inclusive spectra are simple.
More precisely, we assume that bound-state effects can be incorporated
into an effective QCD coupling, which is inserted in the standard soft-gluon
resummation formulas.
We extrapolate therefore the perturbative QCD formulas to a non-perturbative
region by assuming that the relevant non-perturbative effects can be relegated
into an effective coupling.
Since the perturbative formulas involve {\it truncated} expansions in the QCD
coupling, it is clear that our approach is meaningful as long as the
effective coupling remains appreciably smaller than one in all
the relevant integration range.
From Fig.~\ref{coup} we see that our effective coupling is $\approx 0.5$
for a typical soft scale $k_{\perp} \approx 0.5$ GeV
(corresponding to $x_\gamma = 2E_\gamma/m_B \approx 0.9$ in radiative $B$ decays),
i.e. it is reasonably smaller than one.

Since the whole fragmentation process is described in a perturbative
framework, we do not distinguish between the mass of the $B$ meson and
the pole mass of the $b$ quark, i.e. we consistently set $m_b \, = \, m_B$.
We also assume that this effective coupling is {\it universal}, i.e. that it can be
used to describe different processes, and that it can be constructed on the basis
of analyticity arguments.
These are {\it additional} assumptions with respect to the basic one, which could
eventually be relaxed.

Let us remark that the resummed perturbative expansion for
semi-inclusive quantities is incomplete even at the formal level.
For inclusive quantities characterized by a hard scale $Q$, the
cross section can be written in a consistent way as an expansion
in the coupling at the scale $Q$, \beq \label{inclusive}
\sigma_{incl}(Q) \, = \, \sum_{n=0}^{\infty} c_n \,
\alpha_S^n(Q^2), \eeq where the $c_n$'s are numerical coefficients
of order one: no prescription is needed. Semi-inclusive processes
are instead multi-scale processes, characterized by fluctuations
with transverse momenta {\it up} to $Q$; the physical origin is
very clear: a jet with a relatively large invariant mass $m_X$
($\Lambda_{QCD} \ll m_X \ll Q$) can contain very soft partons,
with transverse momenta of the order of the hadronic scale. Unlike
case (\ref{inclusive}), one has to face perturbative contributions
of the form \beq \int_{\approx \, 0}^{Q^2} \frac{ d k_{\perp}^2 }{
k_{\perp}^2 } \, \alpha_S \left( k_{\perp}^2 \right), \eeq where
an ill-defined integration over the Landau pole is made, even for
large $Q \gg \Lambda_{QCD}$. A prescription for the low-energy
behaviour of the coupling is therefore needed in any case. Even if
quark confinement did not exist and partons instead of hadrons
were the asymptotic states, a prescription would  anyway be
necessary to compute resummed cross-sections.

It is clear that our approach has intrinsic and obvious
limitations. The mass of the proton, for example, cannot clearly
be computed by means of perturbative formulas with an effective
coupling inserted in them: a genuinely non-perturbative technique
is mandatory in this case, such as lattice QCD. Our point is that,
with an effective coupling, we want to describe Fermi motion {\it
only}, i.e. a specific non-perturbative effect, not all
non-perturbative effects. We do not aim for example at describing
the $K^*$ peak which appears in the radiative hadron mass
distribution (see Fig.~\ref{rdmx}), or, equivalently, the $\pi$
and $\rho$ peaks which appear in the semileptonic one (see
Fig.~\ref{slmxbab} and Fig.~\ref{slmxbel}). These peaks, occurring
for \beq \label{finalhad} m_X^2 \, \approx \, \Lambda_{QCD}^2,
\eeq are related to final-state hadronization, i.e. to the
recombination of partons into hadrons. This effect has a different
nature with respect to Fermi motion and occurs at a different
scale (cfr. eq.~(\ref{lasolita}) with eq.~(\ref{finalhad})). With
our model, we just want to describe a broad peak in the hadron
mass distribution occurring in region (\ref{lasolita}). A possible
difference between the photon spectra of, let's say, $B \to X_s
\gamma$ and $\Lambda_b \to X_s \gamma$ decays, could not be
described or naturally incorporated in our model, which is a kind
of ``spectator model for spectra''.

The validity of our approach cannot be judged {\it a priori},
but only {\it a posteriori}, by comparing its predictions with experimental data.
One may ask which is the advantage of our approach compared to the standard one
of postulating directly shapes for the non-perturbative components of the spectra
and convolving them with the perturbative ones in the minimal prescription
\cite{Catani:1996yz}.
The answer is that we want to take advantage of the universality properties
of QCD radiation, which are reflected in resummation formulas.
In the standard approach, one has to postulate ad-hoc and un-related shapes for
the non-perturbative components entering different observables,
such as heavy flavor decay spectra, heavy flavor fragmentation, $e^+e^-$ shape variables,
etc.
If universal aspects of QCD dynamics  --- as measured in different processes ---
do exist, such aspects are not easily uncovered with the standard approach.
On the other hand, with our method, such an investigation looks rather natural:
to describe different processes, we use different perturbative formulas
--- quite often the same formulas but with different coefficients ---
with the same effective coupling by assumption and we look at the data
\cite{Dokshitzer:1995qm}.
Our philosophy involves a ``one step'' approach: we deal simultaneously with
perturbative and non-perturbative effects.
The standard method is instead a ``two-step'' approach: one resums the
perturbative long-distance effects in a minimal way --- picking up
just the infrared logarithms
--- and then introduces a physically motivated non-perturbative model.

Another advantage of our approach is that it allows for a simple
extraction of the value of the standard QCD coupling at a
reference scale, f.i.  $\alpha_S\left(m_Z\right)$, by comparing
its predictions with measured $B$ decay spectra. That is because
the model uses ordinary perturbative formulas with a prescription
for the coupling in the low-energy tail, and therefore there is
not any double-counting problem in merging together short-distance
and long-distance effects. A peculiarity of our model is that it
has no free parameters, apart of course the true QCD ones, i.e.
the hadronic scale $\Lambda_{QCD}$ and the quark masses $m_q$'s.
It is therefore ``rigid'', in the sense that there is not a
natural way to tune it to fit the data.

The plan of the paper is as follows.

In sec.~\ref{section1} we summarize the main features of the
ghost-less QCD coupling, which is basically an extrapolation of
the ordinary QCD coupling down to small momentum scales according
to an analyticity principle which removes the Landau pole.

In sec.~\ref{section2} we construct the effective coupling
controlling the evolution of gluon cascades, which are
intrinsically time-like processes.
The absorptive effects related to the decay of the time-like
gluons are included in this effective coupling to all orders
in perturbation theory.

Sec.~\ref{section3} is the main one and describes the model
based on soft gluon resummation in NNLO and on the effective
coupling constructed in the previous section.
A discussion of the relevance of the next-to-next-to-leading-order
effects in our model is also presented.
We also comment on the non-power expansion introduced
in analytic coupling studies.

In sec.~\ref{section4} we apply the model to describe
$B \to X_s \gamma$ decays. We compare its predictions with
the invariant hadron mass distribution measured by BaBar and
with the inclusive photon spectrum measured by Cleo, BaBar and
Belle. Since these spectra are independent from each other,
we obtain for each of them a value of $\alpha_S(m_Z)$
which optimizes the agreement with the data.

In sec.~\ref{section5} we apply the model to the charmless
semileptonic decays $B \to X_u l \nu$.
We compare our predictions with the invariant hadron mass
distribution measured by BaBar and Belle and with the charged
lepton energy spectrum measured by Cleo, BaBar and Belle.
We extract values of $\alpha_S(m_Z)$ as discussed above.

Finally, in sec.~\ref{section6} we draw our conclusions
concerning the agreement of the model with the data.
We also consider natural developments and improvements.

There is also an appendix collecting formulas for the radiative
decay and an appendix with tables of values of the QCD
form factor in our model for a set of values of $\alpha_S(m_Z)$.

\section{Ghost-less Coupling}
\label{section1}

Let us begin considering QCD regularized with an ultra-violet
cut-off $\Lambda_0$ and with a bare coupling $\alpha_0$.
The correlation function
\footnote{
To be accurate, we consider the $q\bar{q}g$
correlation function amputated of all legs and written in terms
of the renormalized fields.
}
representing the quark-gluon interaction
has a perturbative expansion of the form:
\bea
\Gamma_{q\bar{q}g} \left( p_1^2 = p_2^2 = p_3^2 = q^2 \right)
&=&
\alpha_0 \, + \, \beta_0 \, \alpha_0^2 \, \log \frac{\Lambda_0^2}{-q^2-i\varepsilon}
\, + \, \alpha_0^2 \, c
\, + \, \beta_0^2 \, \alpha_0^3 \, \log^2 \frac{\Lambda_0^2}{-q^2-i\varepsilon}
\, + \, \cdots
\nonumber\\
&=&
\frac{\alpha_0}{1 - \beta_0 \, \alpha_0 \, \log \Lambda_0^2/\left( - q^2 - i \varepsilon \right) }
\, + \, \cdots
\eea
where for simplicity's sake we have considered the symmetric point $p_1^2 \, = \, p_2^2 \, = \, p_3^2$.
In the last member we have resummed the well-known geometrical series of the leading logarithms.
$\beta_0 \, = \, \left( 11 \, - \, 2/3 \, n_f \right)/(4\pi)$
is the first-coefficient of the $\beta$-function, $n_f$ is the number of active flavors
and $c$ is a real constant whose explicit expression is not relevant here.
This Green function can be used to define the renormalized QCD coupling \cite{russi}:
\beq
\Gamma_{q\bar{q}g} \left( p_1^2 = p_2^2 = p_3^2 = q^2 \right)
\, \simeq \,
\frac{\alpha_0}{1 - \beta_0 \, \alpha_0 \, \log \Lambda_0^2/\left( - q^2 - i \varepsilon \right) }
\, \equiv \, \alpha(-q^2).
\eeq
To have a {\it real} coupling, one generally assumes a space-like configuration of the
momenta,
\beq
q^2 \, < \, 0
\eeq
and to avoid {\it explicit minus} signs in the renormalization conditions, one defines, like in
Deep-Inelastic-Scattering (DIS):
\footnote{
Note that $Q^2 > 0 $ in the space-like region while $Q^2 < 0$ in the time-like one
and the $\varepsilon$-prescription for $Q^2$ is opposite to that for $q^2$:
\beq
Q^2 \, \equiv \, Q^2 \, - \, i \varepsilon.
\eeq}
\beq
Q^2 \, \equiv \, - \, q^2.
\eeq
We then obtain the usual expression for the renormalized QCD coupling in leading order (LO):
\begin{equation}
\label{start}
\alpha_{lo}(Q^2) \, = \,
\frac{\alpha_0}{1 - \beta_0 \, \alpha_0 \, \log \Lambda_0^2/Q^2 }
\, = \,
\frac{1}{\beta_0 \log Q^2/\Lambda_{QCD}^2},
\end{equation}
where on the last member we have introduced the QCD scale
\beq
\Lambda_{QCD}^2 \, \equiv \, \Lambda_0^2 \, \exp \left[ \, - \, \frac{1}{\beta_0 \, \alpha_0} \right].
\eeq
For notational simplicity, let us write $\Lambda$ in place of $\Lambda_{QCD}$ from now on.
The function on the r.h.s. of eq.~(\ref{start}) has:
\begin{enumerate}
\item
a cut for $Q^2 \, < \, 0$
\footnote{
As usual, the logarithm function is cut along the negative semi-axis, so that:
$\log(-1 \, \pm  \, i \varepsilon) \, = \, \pm \, i \pi$.},
related to the decay of a time-like
gluon into secondary partons,
\beq
\label{gluon_decay}
g^* \, \to \, g \, g, \, q  \, \bar{q}, \, \cdots .
\eeq
This singularity has therefore a clear physical meaning;
\item
a simple pole for $Q^2 \, = \, \Lambda^2$, which does not have any physical
meaning \cite{Aglietti:1995tg}.
This singularity is often called ``Landau ghost'' because of its original
appearance in QED in the interacting electron propagator \cite{landau4}.
It implies a formal divergence of the coupling and a breakdown of the
perturbative scheme.
\end{enumerate}
It has been suggested to replace the usual expression for the coupling
in eq.~(\ref{start}) with a ``ghost-less'' or ``analytic'' coupling
$\bar{\alpha}$ having the following properties \cite{shirkov}:
\begin{enumerate}
\item
it has the same discontinuity along the cut as the standard coupling:
\beq
\label{samedisc}
~~~~~~~~~~~~~~~~~~~
{\rm Disc} \, \bar{\alpha} \, = \, {\rm Disc} \, \alpha
~~~~~~~~~~~~~~~~~~~~~~~~~~~~~~~{\rm (time-like~region);}
\eeq
\item
it is analytic elsewhere in the complex plane.
\end{enumerate}

\begin{figure}[th]
\centerline{\hspace{2cm}\epsfxsize=4truein \epsfbox{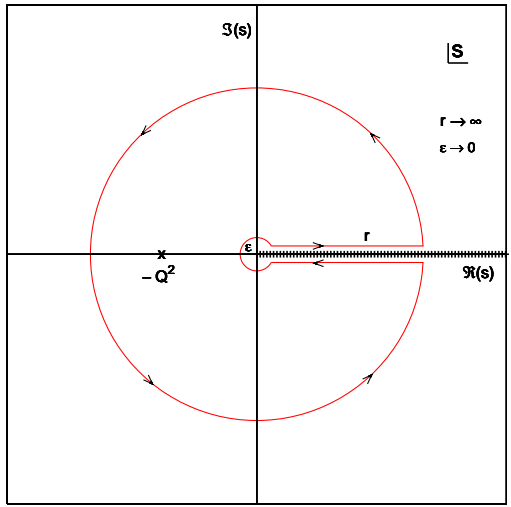}}
\caption{
\small
integration contour $\Gamma$ used to construct the ghost-less coupling.
}
\label{cont}
\end{figure}
Let us now consider the function
\beq
f(s) \, \equiv \, \frac{\bar{\alpha}(-s)}{s \, + \, Q^2},
\eeq
where $Q^2$ is a complex number not lying on the negative axis including
the origin.
By assumption, $f(s)$ is analytic in the complex $s$-plane
cut along the positive axis $s \, \ge \, 0$, except for a simple pole in
\beq
\label{solopolo}
s \, = \, - \, Q^2.
\eeq
We apply the residue theorem to $f(s)$ integrated
along a closed contour $\Gamma$ avoiding the ``physical'' cut for
$s \ge 0$, containing a circle of infinitesimal radius around the origin
$c_{\epsilon}$ ($\epsilon \, \to \, 0$),
a circle at infinity $c_r$ ($r \, \to \, \infty$),
a line above the cut ($s \to s + i\varepsilon$) and
a line below the cut ($s \to s - i\varepsilon$): see Fig.~\ref{cont}.
Being the pole (\ref{solopolo}) the only singularity inside the contour,
we obtain the following expression for the analytic coupling:
\footnote{
Note that, had we taken $Q^2 \le 0$, the pole (\ref{solopolo})
would have been located on the cut and the integral of $f(s)$ over $\Gamma$
would have been zero.}
\bea
\bar{\alpha}(Q^2)
& = &
\frac{1}{2 \pi i}
\oint_{\Gamma}  \frac{\bar{\alpha}(-s)}{s \, + \, Q^2} \, ds
\, = \,
\frac{1}{2 \pi i}
\oint_{ c_{\epsilon} }  \frac{\bar{\alpha}(-s)}{s \, + \, Q^2} \, ds
\, + \,
\frac{1}{2 \pi i}
\oint_{ c_r }  \frac{\bar{\alpha}(-s)}{s \, + \, Q^2} \, ds \, +
\nonumber\\
& & ~~~~~~~~~~~~~~~~~~~~~~~~~
+ \, \frac{1}{2 \pi i}
\int_0^{\infty}  \frac{\bar{\alpha}(- s - i \varepsilon )}{s \, + \, i \varepsilon \, + \, Q^2} \, ds
\, + \,
\frac{1}{2 \pi i}
\int_{\infty}^0 \frac{\bar{\alpha}(- s + i \varepsilon )}{s \, - \, i \varepsilon \, + \, Q^2} \, ds.
\eea
We assume that the contributions of $c_{\epsilon}$ and of $c_r$ vanish.
Since $s + Q^2 \ne 0$ for $s \ge 0$,
\beq
\lim_{\varepsilon \to 0^+} \,
\left[
\frac{\bar{\alpha}(- s - i \varepsilon )}{s \, + \, i \varepsilon \, + \, Q^2}
\, - \,
\frac{\bar{\alpha}(- s + i \varepsilon )}{s \, - \, i \varepsilon \, + \, Q^2}
\right]
\, = \,
\frac{1}{s \, + \, Q^2} \,
\lim_{\varepsilon \to 0^+}
\, \left[ \,
\bar{\alpha}(- s - i \varepsilon )
\, - \,
\bar{\alpha}(- s + i \varepsilon )
\, \right]
\, = \,
\frac{1}{s \, + \, Q^2} \, {\rm Disc}_s \, \bar{\alpha}(-s),
\eeq
where the discontinuity of a function $F(s)$ is defined in general as:
\beq
{\rm Disc}_s F(s) \, \equiv \, \lim_{\varepsilon \, \to \, 0^+}
\left[ \, F(s + i \varepsilon) \, - \, F(s - i \varepsilon) \, \right].
\eeq
Taking into account that for $s \, \ge \, 0$ (see eq.~(\ref{samedisc}))
\beq
{\rm Disc}_s \, \bar{\alpha}(-s)
\, = \, {\rm Disc}_s \, \alpha(-s),
\eeq
we obtain the following integral representation for the ghost-less coupling
in terms of the standard one:
\beq
\label{disp_rel}
\bar{\alpha}(Q^2) \, = \,
\frac{1}{2\pi i}
\int_0^{\infty}  \, \frac{ds}{s \, + \, Q^2} \,
{\rm Disc}_s \, \alpha(-s).
\eeq
Eq.~(\ref{disp_rel}) is just a dispersion relation which, for clarity's sake,
has been fully derived from first principles.
By inserting on the last member the expression for the standard coupling
at lowest order as given by eq.~(\ref{start}), we obtain:
\beq
\bar{\alpha}_{lo}(Q^2)
\, = \, \lim_{\varepsilon \to 0^+} \, \frac{1}{2\pi i \beta_0}
\int_0^{\infty}  \, \frac{ds}{s \, + \, Q^2} \,
\left[
\frac{1}{\log \left(- s/\Lambda^2 - i\varepsilon\right)}
\, - \,
\frac{1}{\log \left(- s/\Lambda^2 + i\varepsilon\right)}
\right].
\eeq
The integral above is elementary. It can also be computed with
the residue theorem by considering the contour $\Gamma$ above.
The circle of infinitesimal radius around the origin and the
circle at infinity give vanishing contributions to the integral.
There are two simple poles in $s = \, - \, Q^2$ and in
$s = \, - \, \Lambda^2$, so that:
\begin{equation}
\label{end}
\bar{\alpha}_{lo}(Q^2) \, = \, \frac{1}{\beta_0}
\left[
\frac{ 1 }{ \log Q^2/\Lambda^2 }
\, - \, \frac{ \Lambda^2 }{ Q^2 - \Lambda^2 }
\right].
\end{equation}
Let us make a few remarks:
\begin{enumerate}
\item
comparing the r.h.s. of eqs.~(\ref{start}) and (\ref{end}),
we see that the ``analyticization'' procedure
had the effect of subtracting the infrared pole in $Q^2 \, = \, \Lambda^2$
by means of a power-suppressed term,
in a minimal way;
\item
the analytic coupling has a constant limit at zero momentum transfer:
\begin{equation}
\lim_{Q^2 \rightarrow 0} \bar{\alpha}_{lo}(Q^2) \, = \, \frac{1}{\beta_0} \, \approx \, O(1);
\end{equation}
\item
the term added to the standard coupling,
\beq
\label{added}
\, - \, \frac{1}{\beta_0}
\, \frac{ \Lambda^2 }{ Q^2 - \Lambda^2 },
\eeq
does not modify the high-energy behavior because it decays as an inverse power of the hard scale,
i.e. infinitely faster than any inverse power of the logarithm of $Q^2$.
In more formal terms, the added term (\ref{added}) is exponentially small in the coupling,
and therefore is always missed in an asymptotic expansion for $Q^2 \, \to \, \infty$:
\begin{equation}
\frac{\Lambda^2}{Q^2 \, - \, \Lambda^2} \, = \,
\frac{ 1 }
{ e^{1/\left[\beta_0\alpha_{lo}(Q^2)\right]} -1 }
\, \approx \,  e^{-1/\left[\beta_0\alpha_{lo}(Q^2)\right]};
\end{equation}
\item
since the power correction has no discontinuity in the time-like region $Q^2 \, < \, 0$,
\beq
{\rm Disc} \, \frac{\Lambda^2}{Q^2 \, - \, \Lambda^2} \, = \, 0
~~~~~~~~~~~~~~~~~~~~~~~{\rm for} ~ Q^2 \, < \, 0,
\eeq
it trivially follows that the analytic coupling has the same discontinuity
as the standard one, as originally requested.
\end{enumerate}
Let us now discuss the extension to next-to-leading order (NLO).
The NLO correction to the standard coupling
\beq
\delta \alpha(Q^2) \, = \, - \, \frac{\beta_1}{\beta_0^3}
\, \log \left( \log \frac{Q^2}{\Lambda^2} \right)
\, \frac{1}{\log^2 Q^2/\Lambda^2},
\eeq
where $\beta_1$ is the second-order coefficient of the $\beta$-function
in the normalization assumed in \cite{noi1},
involves:
\begin{enumerate}
\item
the factor $1/\log^2 Q^2/\Lambda^2$, having
a cut for $Q^2 \, < \, 0$, related to the decay of the time-like
gluon into on-shell partons (see eq.~(\ref{gluon_decay})), and a double pole for
$Q^2 \, = \, \Lambda^2$;
\item
the factor $\log \left( \log Q^2/\Lambda^2 \right)$,
having a cut for $Q^2 \, < \, 0$ related to the ``internal'' logarithm
and another cut for $ 0 \, < \, Q^2 \, < \, \Lambda^2$ related to the
``external'' logarithm.
\end{enumerate}
The singularities for $Q^2 \, = \, \Lambda^2$ and for $0 \, < \, Q^2 \, < \, \Lambda^2$
are unphysical because they refer to the space-like region,
where the virtual gluon cannot decay into physical parton states.
``Analyticization'' can be made as in lowest order:
one requires that the analytic correction term has the same discontinuity
for $Q^2 \, < \, 0$ as the standard one but it is regular elsewhere
in the complex plane:
\begin{equation}
\delta \bar{\alpha}(Q^2)
\, = \,
\frac{1}{2\pi i}
\int_0^{\infty}  \, \frac{ds}{s+Q^2} \,
{\rm Disc}_s \, \delta \alpha(-s).
\end{equation}
The integral above --- unlike the lowest-order case ---
is not elementary but it can easily be made numerically.
The following remarks are in order:
\begin{itemize}
\item
the value of the analytic coupling at zero momentum
transfer is not modified in higher order because:
\begin{equation}
\lim_{Q^2 \, \to \, 0} \delta \bar{\alpha}(Q^2) \, = \, 0;
\end{equation}
\item
it can be shown that $\delta \bar{\alpha}(Q^2)$
has the same logarithmic terms as $\delta \alpha(Q^2)$ \cite{alekseev},
i.e. that the difference resides in power-suppressed terms,
as we have explicitly found for the leading order.
\end{itemize}
The NLO coupling is defined as:
$ \bar{\alpha}_{nl} \, = \, \bar{\alpha}_{lo}
\, + \, \delta \bar{\alpha}.$
Within our accuracy, the next-to-next-to-leading order (NNLO)
corrections to the coupling are also needed:
\begin{equation}
\delta \alpha'(Q^2) \, = \,
\frac{\beta_1^2}{\beta_0^5} \,
\left[
\, \log^2\left( \log \frac{Q^2}{\Lambda^2} \right)
\, - \, \log\left( \log \frac{Q^2}{\Lambda^2} \right)
\, + \, \frac{\beta_0 \, \beta_2 - \beta_1^2}{\beta_1^2}
\right]
\frac{1}{\log^3 Q^2/\Lambda^2},
\end{equation}
where $\beta_2$ is the third-order coefficient of the
$\beta$-function.
One finds similar singularities as in the NLO case, which are removed again according to principle
of ``minimal analyticity'' already used:
\begin{equation}
\delta \bar{\alpha}'(Q^2)
\, = \,
\frac{1}{2\pi i}
\int_0^{\infty} \, \frac{ds}{s+Q^2} \,
{\rm Disc}_s \, \delta \alpha'(-s).
\end{equation}
The NNLO analytic coupling reads (see Fig.~\ref{coup}):
\beq
\bar{\alpha} \, = \, \bar{\alpha}_{lo}
\, + \, \delta \bar{\alpha}
\, + \, \delta \bar{\alpha}'.
\eeq
Let us remark that an expansion in powers of the analytic coupling
$\bar{\alpha}$ is
an asymptotic expansion --- as in the standard case --- because
the logarithmic structure of $\bar{\alpha}$ is the same
as that of $\alpha$. As Fig.~\ref{coup} clearly shows, the standard
coupling and the ghost-less one are barely distinguishable at large
scales \cite{shirkov}.

\section{Effective Coupling for Gluon Cascade}
\label{section2}
As well known from perturbation theory, the emission of a
gluon in a process is accompanied by an additional factor $\alpha$
in the cross section, where $\alpha$ is the tree-level QCD coupling.
In higher orders, one has to consider:
\begin{enumerate}
\item
multiple emissions off the primary color charges ---
the heavy and the light quark in $B$ decays;
\item
secondary emissions off the radiated gluons.
\end{enumerate}
Primary multiple emissions produce the exponentiation
of the one-gluon distribution while secondary emissions
produce the decay of the radiated gluons
into secondary partons --- see eq.~(\ref{gluon_decay}).
In the case of form factors, which are inclusive with respect to
gluon decays, these higher-order terms have the main
effect of replacing the tree-level coupling with an effective
coupling evaluated at the transverse momentum of the primary emitted
gluon \cite{Amati:1980ch}:
\beq
\alpha \, \to \,
\tilde{\alpha}(k_{\perp}^2),
\eeq
where
\begin{equation}
\label{deftime}
\tilde{\alpha}(k_{\perp}^2) \, \equiv \, \frac{i}{2 \pi} \, \int_0^{k_{\perp}^2} \, d s
\, {\rm Disc}_s \, \frac{ \alpha(-s) }{ s }.
\end{equation}
The coupling $\tilde{\alpha}(k_{\perp}^2)$ is characteristic of the QCD cascade and
it is given by the integral
of the discontinuity of the (interacting) gluon propagator over virtualities $s$
cut-off by the primary gluon transverse momentum.
Let us remark that the cascade (or effective or time-like) coupling always refers
to time-like kinematics.

\begin{figure}[th]
\centerline{\epsfxsize=4truein \epsfbox{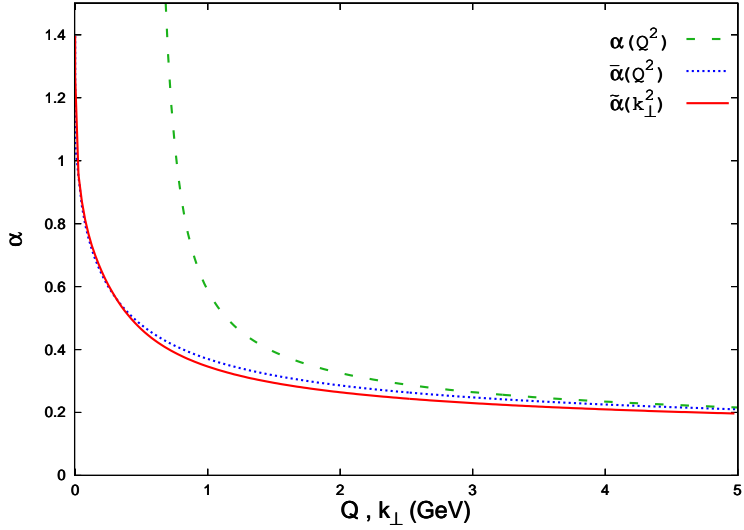}}
\caption{
\small
QCD couplings in NNLO for a fixed number of active flavors $n_f \, = \, 3$ and
$\Lambda_{QCD}^{(3)} \, = \, 0.7$ GeV.
Dashed line (green): standard coupling $\alpha(Q^2)$; dotted line (blue): ghost-less or analytic coupling
$\bar{\alpha}(Q^2)$; continuous line (red): cascade or time-like coupling $\tilde{\alpha}(k_{\perp}^2)$.
}
\label{coup}
\end{figure}

The prescription at the root of our model is simply to replace the standard coupling on the r.h.s. of
eq.~(\ref{deftime}) with the ghost-less coupling constructed in the previous section:
\begin{equation}
\label{deftime2}
\tilde{\alpha}(k_{\perp}^2) \, = \, \frac{i}{2 \pi} \, \int_0^{k_{\perp}^2} \, d s
\, {\rm Disc}_s \, \frac{ \bar{\alpha}(- s) }{ s }.
\end{equation}
If we neglect the $-i\pi$ terms in the integral over the discontinuity
--- i.e. the absorptive effects --- the cascade coupling exactly
reduces  to the ghost-less one:
\beq
\tilde{\alpha}(k_{\perp}^2) \, \to \, \bar{\alpha}(k_{\perp}^2).
\eeq
To render our model as accurate as possible, we include such absorptive
effects and perform the integral on the r.h.s. of eq.~(\ref{deftime2}) exactly.
By inserting the analytic coupling at LO in the integrand on the
r.h.s. of eq.~(\ref{deftime2}), we obtain for the effective coupling:
\begin{equation}
\label{dopo}
\tilde{\alpha}_{lo}(k_{\perp}^2) \, = \,
\frac{1}{2\pi i \beta_0}
\left[
        \log\left( \log \frac{k_{\perp}^2}{\Lambda^2} + i \pi \right)
\, - \, \log\left( \log \frac{k_{\perp}^2}{\Lambda^2} - i \pi \right)
\right].
\end{equation}
At NLO, one has to add the contribution:
\begin{eqnarray}
\delta \tilde{\alpha}(k_{\perp}^2) &=&
 \, \frac{\beta_1}{2\pi i \, \beta_0^3} \,
\Bigg[
        \frac{ \log\Big( \log k_{\perp}^2/\Lambda^2 + i \pi \Big) + 1 }
             { \log k_{\perp}^2/\Lambda^2 + i \pi }
\, - \, \frac{ \log\Big( \log k_{\perp}^2/\Lambda^2 - i \pi \Big) + 1 }
             { \log k_{\perp}^2/\Lambda^2 - i \pi }
\Bigg].
\end{eqnarray}
The NNLO corrections read:
\begin{eqnarray}
\delta \tilde{\alpha}'(k_{\perp}^2) &=& - \, \frac{\beta_1^2}{4\pi i \beta_0^5}
\left[
     \frac{ \log^2\left( \log k_{\perp}^2/\Lambda^2 + i \pi \right) }
             { \left( \log k_{\perp}^2/\Lambda^2 + i \pi \right)^2 }
\, - \, \frac{ \log^2\left( \log k_{\perp}^2/\Lambda^2 - i \pi \right) }
             { \left( \log k_{\perp}^2/\Lambda^2 - i \pi \right)^2 }
\right] \, + \,
\nonumber\\
&& + \, \frac{ \beta_1^2 - \beta_0 \, \beta_2 }{4\pi i \beta_0^5}
\left[
        \frac{ 1 }{ \left( \log k_{\perp}^2/\Lambda^2 + i \pi \right)^2 }
\, - \, \frac{ 1 }{ \left( \log k_{\perp}^2/\Lambda^2 - i \pi \right)^2 }
\right].
\end{eqnarray}
The time-like coupling in NNLO is simply the sum of the above terms:
\beq
\tilde{\alpha}(k_{\perp}^2) \, = \,
 \tilde{\alpha}_{lo}(k_{\perp}^2) \, + \,
\delta \tilde{\alpha}(k_{\perp}^2) \, + \,
\delta \tilde{\alpha}'(k_{\perp}^2).
\eeq
Let us make a few remarks:
\begin{enumerate}
\item
the cascade coupling is very close to the ghost-less one for very small scales,
let's say less than $1$ GeV (see Fig.~\ref{coup}).
That is partly a consequence of the fact that both couplings
have the same limit at zero momentum, $1/\beta_0$, and is
partly accidental \cite{shirkov}.
The cascade coupling is instead smaller than the standard coupling or the
ghost-less one in the perturbative region, at large scales, because
it has an additional negative third-order contribution
$\approx \, - \, 1/\log^3 Q^2$ --- see next point;
\item
the time-like coupling $\tilde{\alpha}$ has an expansion in powers of the standard $\overline{MS}$
coupling $\alpha$ of the form:
\begin{equation}
\label{coup_scheme}
\tilde{\alpha} \,  = \, \alpha \, - \, \frac{\left(\pi\beta_0\right)^2}{3} \, \alpha^3
\, - \, \frac{5}{6} \, \frac{\beta_1}{\beta_0} \, \left(\pi\beta_0\right)^2 \, \alpha^4
\, + \, O(\alpha^5).
\end{equation}
The relation above can be considered as an ordinary change of scheme for the coupling
starting at third order;
\item
the $\tilde{\beta}$ function for the time-like coupling, defined by the relation
\beq
\frac{ d \, \tilde{\alpha} }{d \log k_{\perp}^2} \, = \, \tilde{\beta}(\tilde{\alpha})
\, =
\, - \, \tilde{\beta}_0 \, \tilde{\alpha}^2
\, - \, \tilde{\beta}_1 \, \tilde{\alpha}^3
\, - \, \tilde{\beta}_2 \, \tilde{\alpha}^4
\, - \, \cdots,
\eeq
has a negative third-order
coefficient\footnote{
The first two coefficients are, as well known, invariant under
a change of scheme: $\tilde{\beta}_0 \, = \, \beta_0$, $\tilde{\beta}_1 \, = \, \beta_1$.
},
\begin{equation}
\tilde{\beta}_2 \, = \,
\beta_2 \, - \, \frac{1}{3} \, \left(\pi\beta_0\right)^2 \beta_0 \, < \, 0,
\end{equation}
in agreement with the fact that the coupling saturates at small scales;
\item
An expansion in powers of $\tilde{\alpha}(k_{\perp}^2)$ is not an asymptotic
expansion for $k_{\perp}^2 \, \to \, \infty$ because $\tilde{\alpha}$ even
at LO contains infinitely many inverse powers of $\log k_{\perp}^2 $.
\end{enumerate}
The decoupling relations for the time-like coupling differ from the
ones for the standard $\overline{MS}$ coupling and read:
\beq
\tilde{\alpha}_{n_f} \, = \, \tilde{\alpha}_{n_f-1}
\, - \, \left( \frac{11}{72 \pi ^2} \, - \, \frac{17}{54} \, + \, \frac{n_f}{54} \right)
\, \tilde{\alpha}_{n_f-1}^{\,3},
\eeq
where $\tilde{\alpha}_{n_f}$ is the time-like coupling with $n_f$ active flavors
and $\tilde{\alpha}_{n_f-1}$ with $n_f-1$.
The above relation has to be imposed at a scale $\mu$ such that
$\bar{m}(\mu) \, = \, \mu$, where $\bar{m}(\mu)$ is the
$\overline{MS}$ running mass of the decoupling quark.

Let us end this section summarizing the basic steps taken in the construction of the
effective coupling for the gluon cascade of our model:
\begin{enumerate}
\item
subtraction of the Landau pole from the ordinary QCD coupling;
\item
inclusion of the absorptive effects related to the decay of
time-like gluons in the coupling controlling jet evolution.
\end{enumerate}

\subsection{Coupling in the DMW model}

Let us now evaluate the quantity
\beq
\alpha_0 \, \equiv \, \frac{1}{\mu_I} \int_0^{\mu_I} d k_{\perp} \tilde{\alpha}\left( k_{\perp}^2 \right)
\eeq
parameterizing the leading non-perturbative effects in the Dokshitzer--Marchesini--Webber (DMW) model
\cite{Dokshitzer:1995qm}.
With $\mu_I \, = \, 2$ GeV we obtain in our model
$\alpha_0 \, = \, 0.40$ for $\alpha(m_Z) \, = \, 0.12$ and
$\alpha_0 \, = \, 0.44$ for $\alpha(m_Z) \, = \, 0.125$.
In general, we find that $\alpha_0$ is roughly linear in $\alpha(m_Z)$.
A fit to $e^+e^-$ shape variables data using next-to-leading resummed
formulas gives $\alpha_0 \simeq 0.45$
\footnote{
G.~Salam: private communication.
}.

\section{Threshold Resummation with Effective Coupling}
\label{section3}

In this section we describe a model for threshold resummation
in semi-inclusive beauty decays based on the effective coupling considered
in the previous section.
Basically, we replace in the resummation exponent the standard coupling
with the effective one.

\subsection{$N$-space}

In order to factorize multiple soft-gluon kinematic constraints,
a transformation to $N$-space is required:
\begin{equation}
\sigma_N(\alpha) \, = \, \int_0^1 \, (1-t)^{N-1} \, \sigma(t;\,\alpha) \, dt,
\end{equation}
where
\begin{equation}
\sigma(t;\,\alpha) \, = \, \delta(t) \, - \, \frac{C_F \,\alpha}{\pi}
\left(\frac{\log t}{t}\right)_+
                \, - \, \frac{ 7 \, C_F\, \alpha}{4\,\pi} \left(\frac{1}{t}\right)_+
                \, + \, O(\alpha^2),
\end{equation}
is the differential QCD form factor in the notation of \cite{noi1}.
$C_F \, = \, (N_C^2-1)/(2N_C) \, = \, 4/3$ with $N_C = 3$ the number of colors
and the plus distributions are defined as usual as:
\begin{equation}
P(t)_+ \, \equiv \, P(t) \, - \, \delta(t) \int_0^1 P(t')  \, d t'.
\end{equation}
The form factor has an exponential form in $N$-space:
\begin{equation}
\sigma_N(\alpha) \, = \, e^{ G_N(\alpha) },
\end{equation}
where the exponent of the form factor reads:
\begin{equation}
\label{expoGN}
G_N(\alpha) \, = \,
\int_0^1 \frac{dy}{y} \left[ (1-y)^{N-1} - 1 \right]
\left\{
\int_{Q^2 y^2}^{Q^2 y} \frac{dk_{\perp}^2}{k_{\perp}^2}
\tilde{A}\left[\tilde{\alpha}(k_{\perp}^2)\right]
\, + \, \tilde{B}\left[\tilde{\alpha}(Q^2 y)\right]
\, + \, \tilde{D}\left[\tilde{\alpha}(Q^2 y^2)\right]
\right\},
\end{equation}
with $Q \, = \, w \, m_B$ being the hard scale and $w \, \equiv \, 2E_X/m_B$.
The functions $\tilde{A}(\tilde{\alpha}),\, \tilde{B}(\tilde{\alpha})$
and $\tilde{D}(\tilde{\alpha})$
have expansions in powers of the effective coupling:
\begin{eqnarray}
\tilde{A}(\tilde{\alpha}) &=& \sum_{n=1}^{\infty} \tilde{A}_n \, \tilde{\alpha}^n
\, = \, \tilde{A}_1 \, \tilde{\alpha} \, + \, \tilde{A}_2 \, \tilde{\alpha}^2
\, + \,\tilde{A}_3 \, \tilde{\alpha}^3 \, + \, \cdots;
\\
\tilde{B}(\tilde{\alpha}) &=& \sum_{n=1}^{\infty} \tilde{B}_n \, \tilde{\alpha}^n
\, = \, \tilde{B}_1 \, \tilde{\alpha} \, + \, \tilde{B}_2 \, \tilde{\alpha}^2
\, + \, \cdots;
\\
\tilde{D}(\tilde{\alpha}) &=& \sum_{n=1}^{\infty} \tilde{D}_n \, \tilde{\alpha}^n
\, = \, \tilde{D}_1 \, \tilde{\alpha} \, + \, \tilde{D}_2 \, \tilde{\alpha}^2 \, + \, \cdots.
\end{eqnarray}
The resummation constants for the cascade coupling are obtained
from the standard ones (usually in the $\overline{MS}$ scheme)
by imposing equalities such as:
\begin{equation}
\tilde{A}(\tilde{\alpha}) \, = \, A(\alpha),
\end{equation}
where
\begin{equation}
A(\alpha) \, = \, \sum_{n=1}^{\infty} A_n \, \alpha^n \, = \,
A_1 \, \alpha  \, + \, A_2 \, \alpha^2 \, + \, A_3 \, \alpha^3 \, + \cdots
\end{equation}
is the standard double-logarithmic function
\footnote{A compilation of the resummation constants in our normalization,
with references to the original papers, can be found in \cite{noi1}.}.
Expressing the cascade coupling in terms of the standard one,
according to eq.~(\ref{coup_scheme}), we obtain:
\footnote{
Analogous relations hold for the $\tilde{B}_i$'s and the $\tilde{D}_i$'s.}
\begin{eqnarray}
\tilde{A}_1 &=& A_1 ;
\\
\tilde{A}_2 &=& A_2 ;
\\
\tilde{A}_3 &=& A_3 \, + \, \frac{\left(\pi\beta_0\right)^2}{3} \, A_1;
\\
\tilde{A}_4 &=& A_4 \, + \, \frac{2}{3} \, \left(\pi\beta_0\right)^2 \, A_2
\, + \,  \frac{5}{6} \, \frac{\beta_1}{\beta_0} \, \left(\pi\beta_0\right)^2 \, A_1.
\end{eqnarray}
The first two coefficients $A_1$ and $A_2$ are the same for both couplings $\alpha$ and
$\tilde{\alpha}$, while the third-order one $A_3$
is modified going to the time-like coupling by a contribution proportional to the first-order
coefficient. For $n_f \, = \, 3$, $\tilde{A}_3 \approx 1$ is larger than $A_3 \approx 0.3$
by a factor 3, but it is still acceptably small.

We have found that the inclusion of the NNLO terms involving $\tilde{A}_3$, $B_2$ and $D_2$ ---
in particular $\tilde{A}_3$ -- is crucial for a good description of the experimental data.
The NLO spectra are indeed peaked at too low hadron invariant masses and a sizable and positive
value for $\tilde{A}_3$ suppresses the elastic region and shifts the spectra to higher $m_X$'s.
The model could be improved by including NNNLO terms, which require the
knowledge of the coefficients $A_4$, $B_3$ and $D_3$; at present, only $B_3$ is
analytically known \cite{Moch:2005ba}.

Our model has been constructed by means of a power expansion in a single (effective) coupling
$\tilde{\alpha}$, i.e. higher orders are proportional to $\tilde{\alpha}^n$.
In \cite{shirkov} and in \cite{Aglietti:2004fz} a non-power expansion had been proposed
involving a different coupling for any $n$, which has interesting theoretical properties.
In second order ($n = 2$), for example, one has the coupling
\beq
\widetilde{\alpha^2}( k_{\perp}^2 ) \, = \,
\frac{1}{\beta_0^2 \left( \pi^2 \, + \, \log^2 k_{\perp}^2/\Lambda^2 \right)}
\eeq
in place of $\tilde{\alpha}( k_{\perp}^2 )^2$, with $\tilde{\alpha}( k_{\perp}^2 )$
given by eq.~(\ref{dopo}).
We have found that the non-power expansion does not offer a good description of the
measured spectra.
That is because
\beq
\widetilde{\alpha^2}( k_{\perp}^2 ) \, \to \, 0
~~~~~{\rm while}~~~~~
\tilde{\alpha}( k_{\perp}^2 )^2 \, \to \, \frac{1}{\beta_0^2} \, \approx \, O(1)
~~~~~{\rm for}~~~~~
k_{\perp}^2 \, \to \, 0.
\eeq
That implies that second-order effects are suppressed in the soft region with the non-power
expansion compared to the power expansion case.
In general, the non-power expansion renders the higher-order effects very small \cite{shirkov}.
But, as discussed above, in beauty decays, sizable third-order effects are needed to take
the theoretical curves close to the data, disfavoring the non-power expansion.

In order to include as many corrections as possible
--- higher order $\log N$ terms, $1/N$ contributions, etc. ---
in agreement with the philosophy described in the introduction,
we make the integration over $y$ in $G_N$ exactly, in numerical way.
This is possible because the time-like coupling $\tilde{\alpha}(k_{\perp}^2)$, unlike the
standard one, does not have the Landau singularity and is regular
for any $k_{\perp}^2 \, \ge \, 0$.

\subsection{Inverse Transform}

The form factor in momentum space is obtained by inverse
transform:
\begin{equation}
\label{inverse}
\sigma(t; \, \alpha)
\, = \,
\int_{C-i\infty}^{C+i\infty} \frac{dN}{2\pi i}
(1-t)^{-N} \, \sigma_N(\alpha),
\end{equation}
where the constant $C$ is chosen so that the integration contour
in the $N$-plane lies to the right of all the singularities of $\sigma_N(\alpha)$.
In order to correctly implement multi-parton kinematics, the inverse transform
from $N$-space back to $x$-space is also made exactly in numerical way.
Let us note that no prescription --- such as the minimal prescription in the standard formalism
\cite{Catani:1996yz} --- is needed in our model because $\sigma_N(\alpha)$ is analytic
for ${\rm Re} \, N \, > \, 0$.

\section{Radiative Decay}
\label{section4}

The event fraction or partially-integrated rate for the radiative $B$ decay
\beq
B \, \to \, X_s \, + \, \gamma
\eeq
can be written as \cite{me,acg}:
\begin{equation}
\frac{ 1 }{ \Gamma^{(0)}_r } \int_0^t\frac{d\Gamma_r}{dt'} dt' \, = \,
K_r(\alpha) \, \Sigma(t;\alpha) \, + \, D_r(t;\,\alpha)
\end{equation}
where
\beq
t \, \equiv \, \frac{m_X^2}{m_B^2}
\eeq
is a dimensionless variable,
\begin{equation}
\Sigma(t;\,\alpha) \, = \, \int_0^t \sigma(t';\,\alpha) \, dt'
 \, = \, 1 \, - \, \frac{C_F \,\alpha}{2\pi} \, \log^2 t
                \, - \, \frac{ 7 \, C_F\, \alpha}{4\,\pi} \, \log t
                \, + \, O(\alpha^2)
\end{equation}
is the partially-integrated form factor and
\begin{equation}
\Gamma^{(0)}_r \, = \, \frac{\alpha_{em}}{\pi}
\frac{G_F^2 \, m_b^3 \, \bar{m}_b^2 } {32 \pi^3} \, C_7^2
\end{equation}
is the lowest-order inclusive width.
$m_b \approx 5$ GeV is the beauty pole mass while $\bar{m}_b$ is the $\overline{MS}$ mass
evaluated in $\mu\,=\,m_b$. Their relation reads:
\begin{equation}
\bar{m}_b \, = \,
\left[ 1 - \frac{\alpha(m_b) C_F}{\pi} + O(\alpha^2) \right] m_b \, \simeq \, 0.9 \, m_b.
\end{equation}
$K_r(\alpha)$ is a short-distance coefficient function specific for this process
and having an expansion in powers of $\alpha$:
\beq
K_r(\alpha) \, = \, 1 \, + \, \alpha \, K_r^{(1)} \, + \, \alpha^2 \, K_r^{(2)}
\, + \, O(\alpha^3).
\eeq
The explicit expression of the first-order term reads:
\beq
K_r^{(1)} \, = \, \frac{1}{2\pi} \sum_{i=1}^8 \frac{C_i}{C_7} \, {\rm Re} \, r_i,
\eeq
where the $C_i$'s are short-distance coefficient functions entering the effective $b\to s \gamma$
Hamiltonian, ${\cal H}^{b\to s \gamma}$, whose numerical values are given in the appendix,
and the $r_i$'s are complex constants.
$D_r(t;\,\alpha)$ is a process-dependent remainder function, which is included to correctly
describe also the high jet mass region $t \, \approx \, O(1)$.
In our leading-twist analysis, this function can be computed in perturbation theory
and starts in first order:
\beq
D_r(t;\,\alpha) \, = \, \alpha \, D_r^{(1)}(t) \, + \, \alpha^2 \, D_r^{(2)}(t)
\, + \, O(\alpha^3),
\eeq
with
\beq
D_r^{(1)}(t) \, = \, \frac{1}{ \pi }
\sum_{i\le j}^{1,8} \frac{C_i \, C_j}{ C_7^2 } \, f_{ij}^{(1)}(t).
\eeq
The $f_{ij}(t)$'s are functions whose explicit expression are given in the appendix.

Since the total width $\Gamma_r$ is infrared divergent beyond tree level,
because of soft photon effects occurring in the spectrum for $t \rightarrow 1$,
it is convenient to define an event fraction normalized to the partial rate
\begin{equation}
\Gamma_r(\delta) \, \equiv \, \int_0^{\delta} \frac{d\Gamma_r}{dt'} \, dt',
\end{equation}
where $\delta <1$ is a parameter. That is also convenient for experimental reasons:
due to large backgrounds, the presently accessible range of hadron masses is
at the most $0 < t < 0.3$ (see later).
The event fraction normalized to $\Gamma_r(\delta)$ reads:
\begin{equation}
R_{\delta}(t) \,= \, \frac{1}{\Gamma_r(\delta)} \, \int_0^t \, \frac{d\Gamma_r}{dt'} \, dt'.
\end{equation}
For $\delta\rightarrow 1$, $R_{\delta}(t)$ tends to the standard event fraction $R_r(t)$.
The normalization condition is:
\begin{equation}
\label{normalization2}
R_{\delta}(\delta) \,= \, 1.
\end{equation}
The differential spectrum is obtained by differentiation:
\begin{equation}
\frac{d\Gamma_r}{dt} \,  = \, \Gamma_r(\delta) \, \frac{d R_{\delta}}{dt}.
\end{equation}
A resummed expression of the following form holds:
\begin{equation}
R_{\delta}(t; \, \alpha) \,= \, C_{\delta}(\alpha) \,\Sigma_{\delta}(t; \, \alpha)
\, + \, D_{\delta}(t; \, \alpha),
\end{equation}
where we have defined the form factor
\begin{equation}
\Sigma_{\delta}(t; \, \alpha) \, \equiv \,
\frac{ \Sigma(t; \, \alpha) }{ \Sigma(\delta; \, \alpha) },
\end{equation}
which is normalized as
\begin{equation}
\Sigma_{\delta}(\delta; \, \alpha) \, = \, 1.
\end{equation}
The normalization condition (\ref{normalization2}) gives:
\begin{equation}
C_{\delta}(\alpha) \, + \, \, D_{\delta}(\delta; \,\alpha) \, = \, 1.
\end{equation}
The expansions of the coefficient function and the remainder function read:
\begin{eqnarray}
C_{\delta}(\alpha) &=& 1 \, + \, \alpha \, C_{\delta}^{(1)}
+ \, \alpha^2 \, C_{\delta}^{(2)} \, + \, O(\alpha^3) ;
\\
D_{\delta}(t;\,\alpha) &=& \alpha \, D_{\delta}^{(1)}(t) \, + \,
        \alpha^2 \, D_{\delta}^{(2)}(t) \, + \, O(\alpha^3),
\end{eqnarray}
with
\begin{equation}
C_{\delta}^{(1)} \, = \,  - \, \frac{ D_r^{(1)}(\delta) } {\Sigma(\delta;\,\alpha)};
~~~~~~~~~~
D_{\delta}^{(1)}(t) \, = \, \frac{ D_r^{(1)}(t) }{ \Sigma(\delta;\,\alpha) }.
\end{equation}
Note that we only expand $D_r(t;\,\alpha)$ in powers of $\alpha$
and not $\Sigma(\delta;\,\alpha)$, because, for sufficient small $\delta$,
one can have $\alpha \log^2\delta \, \approx \, O(1)$,
implying need for resummation to any order in $\alpha$.
For $\delta \, = \, 0.26$, one obtains $C_{\delta}^{(1)}\, \simeq \, - \, 0.48$,
i.e. a $O(10\%)$ correction to the tree-level coefficient function
$(\Sigma(t = 0.26; \,\alpha) \, = \, 1.10)$.

\subsection{Phenomenology}

\begin{figure}[th]
\centerline{\epsfxsize=4truein \epsfbox{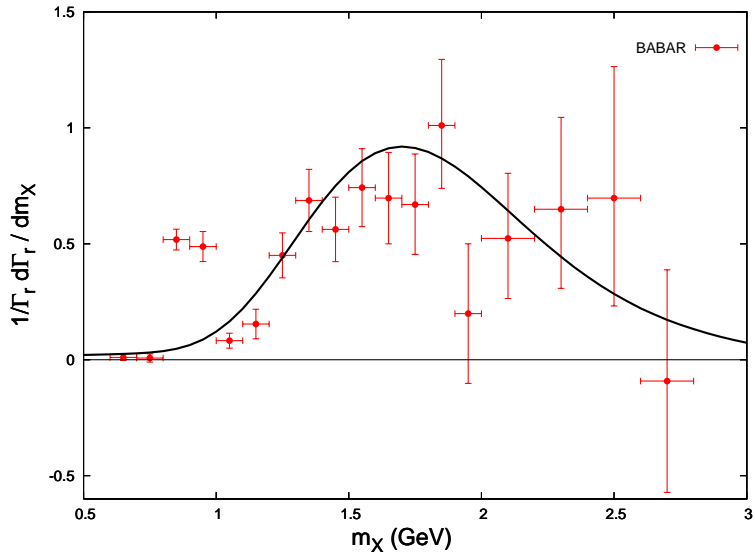}}
\caption{
\small
$B\to X_s \gamma$ invariant hadron mass distribution from BaBar compared to our model
for $\alpha_S(m_Z)\, = \, 0.123$.
The theory and the data are normalized to one in the experimentally accessible region.
}
\label{rdmx}
\end{figure}

In Fig.~\ref{rdmx} we compare the invariant hadron mass distribution
for the radiative decay, $d\Gamma_r/d m_X$, obtained with our model with
experimental data from the BaBar collaboration \cite{babargam2}.
The data show a rather pronounced $K^{*}$ peak, which clearly cannot be accounted for
in a perturbative QCD framework
\footnote{
To have a point-to-point description of the data, one has to include by hand the contribution of
this resonance, by means of one or more free parameters.}.
We have therefore discarded in the analysis the data points with $m_X \, < \, 1.1$ GeV
\footnote{
$m_{K^*} \, = \, 892 $ MeV and $\Gamma_{K^*} \, = \, 51$ MeV.
}.
We obtain a minimum $\chi^2 \, = \, 12 $ for $\alpha_S(m_Z) \, = \, 0.1255$
for $13$ data points, i.e. for $12$ degrees of freedom (d.o.f.) because of the fixed normalization.
Since we deal with the standard QCD coupling, let us write $\alpha_S$ from this section till
the end of the paper.
To improve the agreement of the theory with the data and to estimate the error on $\alpha_S(m_Z)$,
we have performed a Gaussian smearing of $\Delta m_X \, = \, 300$ MeV of the data points
{\it and} of the theoretical distribution, and we have discarded the points
with $m_X \, < \, 800$ MeV.
We obtain a minimum $\chi^2 \, = \, 6.8$ for $\alpha_S(m_Z) \, = \, 0.1205$
for $15$ d.o.f..
By taking as an estimate of $\alpha_S(m_Z)$ the average of the above values
and as an estimate of the error their difference, we quote:
\footnote{
We have taken $\bar{m}_b \, = \, 4.8$ GeV, $\bar{m}_c \, = \, 1.4$ GeV and
$\bar{m}_s \, = \, 0.3$ GeV in the decoupling relations.
In general, changing the $\overline{MS}$ masses in a reasonable range
modifies the theoretical predictions in a negligible way.
Increasing the $\overline{MS}$ masses is roughly equivalent to a slight increase of
$\alpha_S(m_Z)$. That is because, lowering the renormalization scale, the QCD coupling
rises faster for a smaller number of active flavors.}
\beq
~~~~~~~~~~~
\alpha_S(m_Z) \, = \, 0.123 \, \pm \, 0.003
~~~~~~~~~~~~~~~~~~~~~~ \left( m_{Xs}: \, {\rm BABAR} \right).
\eeq

\begin{figure}[th]
\centerline{\epsfxsize=4truein \epsfbox{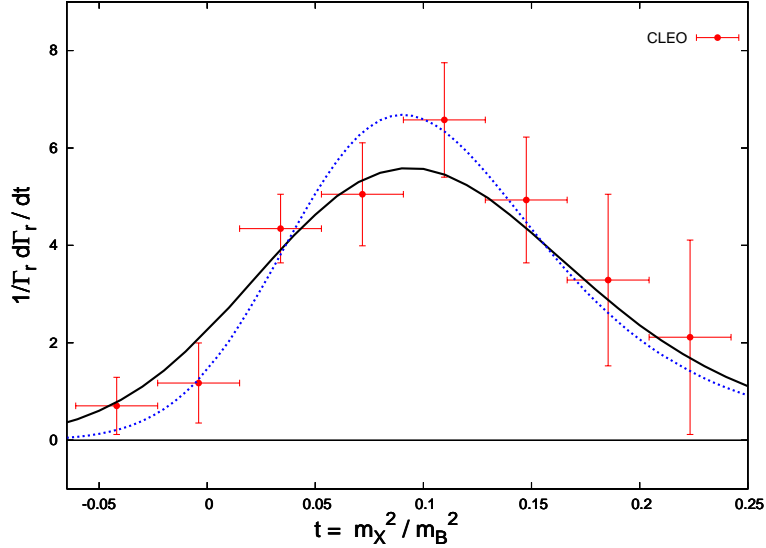}}
\caption{
\small
$B\to X_s \gamma$ photon spectrum from Cleo compared to our model.
Dotted line (blue): $\alpha_S(m_Z) \, = \, 0.118$ and $\sigma_\gamma \, = \, 100$ MeV
to model the Doppler effect (see text);
continuous line (black): $\alpha_S(m_Z) \, = \, 0.117$ and $\sigma_\gamma \, = \, 150$ MeV.}
\label{efcleo}
\end{figure}

In Fig.~\ref{efcleo} we compare the photon energy spectrum computed in the framework of
our model with data from the Cleo Collaboration \cite{cleogam1}.
In the $B$ rest-frame,
\beq
t \, = \, 1 \, - \, \frac{2E_{\gamma}}{m_B}.
\eeq
The photon energies are however measured in the $\Upsilon(4S)$ rest frame, in which
the $B$ mesons have a small, non-relativistic motion.
In order to model the Doppler effect, we have convoluted
the theoretical curve for $E_\gamma$ --- computed with a $B$ meson at rest --- with a
normal distribution of $\sigma_\gamma \, = \, 150$ MeV, as suggested by Cleo itself.
Let us note that the Doppler effect is sufficient to completely wash out
the $K^*$ peak.
We obtain a minimum $\chi^2 \, = \, 3.8$ for $\alpha_S(m_Z) \, = \, 0.117$
for $7$ d.o.f..
Assuming complete independence of the experimental points, we allow the $\chi^2$
to raise by one unit to estimate the error and we obtain:
\beq
~~~~~~~~~
\alpha_S(m_Z) \, = \, 0.117 \, \pm \, 0.004
~~~~~~~~~~~~~~~~~~~~~ \left( E_{\gamma}: \, {\rm CLEO},~ \sigma_\gamma \, = \, 150~{\rm MeV} \right).
\eeq
To check the modeling above of the Doppler effect, we have used the following method.
We have converted the $m_{Xs}$ distribution by BaBar above to a photon
spectrum in the $B$ rest-frame and we have convoluted it with a normal distribution
with a variable $\sigma_\gamma$, obtaining the points $(x_i, y_i'(\sigma_\gamma), \sigma_i')$.
We have then minimized the quantity
\beq
\label{idea}
H\left( \sigma_\gamma \right) \, \equiv \,
\sum_i \frac{ \left[ y_i-y_i'(\sigma_\gamma) \right]^2 }{ \sigma_i^2 + \sigma_i'^{\,2} }
\eeq
with respect to $\sigma_\gamma$, where $(x_i, y_i, \sigma_i)$ are the Cleo data
\footnote{
Let us remark however that the two spectra entering eq.~(\ref{idea}) are independent
on each other.
}.
We have found a minimum of $H(\sigma_\gamma)$ for $\sigma_\gamma  \simeq  100$~MeV,
which gives similar results to the analysis with $\sigma_\gamma \, = \, 150$ MeV:
\beq
\alpha_S(m_Z) \, = \, 0.118 \, \pm \, 0.003
~~~~~~~~~~~~~~~~ \left( E_{\gamma}: \, {\rm CLEO},~\sigma_\gamma \, = \, 100~{\rm MeV} \right),
\eeq
with $\chi^2 \, = \, 3.4$.
As intuitively expected, reducing $\sigma_\gamma$ produces a shaper spectrum.

\begin{figure}[th]
\centerline{\epsfxsize=4truein \epsfbox{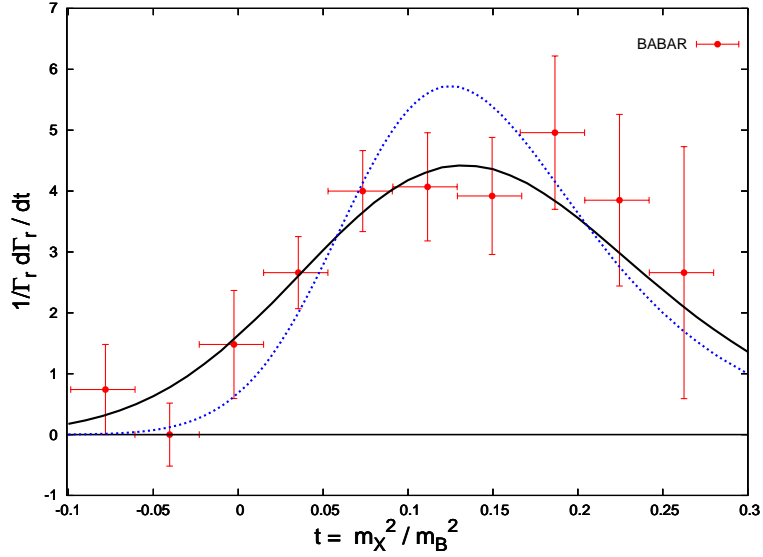}}
\caption{
\small
$B\to X_s \gamma$ photon spectrum from BaBar compared to our model.
Dotted line (blue): $\alpha_S(m_Z) \, = \, 0.130$ and $\sigma_\gamma \, = \, 100$ MeV;
continuous line (black): $\alpha_S(m_Z) \, = \, 0.129$ and  $\sigma_\gamma \, = \, 200$ MeV.
}
\label{efbabar}
\end{figure}

In Fig.~\ref{efbabar} we compare the predictions of our model with a spectrum
from the BaBar collaboration \cite{babargam1}.
The BaBar spectrum is somewhat softer than the Cleo one --- even
though the difference is within one standard deviation; we have interpreted
this difference as a resolution effect and we have convoluted our
theoretical curve with a normal distribution with a slightly larger
standard deviation, $\sigma_\gamma \, = \, 200$ MeV
\footnote{A more sophisticated analysis from the experimentalists,
including the true resolution functions, is strongly encouraged!}.
We obtain a minimum $\chi^2 \, = \, 5.1$ for $9$ d.o.f..
Performing a similar analysis as for the Cleo data, we obtain:
\beq
~~~~~~~~~~~~~~~~~~~~~~~~~
\alpha_S(m_Z) \, = \, 0.129 \, \pm \, 0.005
~~~~~~~~~~~~~~~~~~~~~~~~~
\left( E_{\gamma}: \,{\rm BABAR},~\sigma_\gamma \, = \, 200~{\rm MeV}\right).
\eeq
Following the minimization procedure above for the Cleo spectrum
(see eq.~(\ref{idea})), we obtain also for the BaBar photon spectrum
$\sigma_\gamma \, \simeq \, 100$ MeV, to give:
\beq
~~~~~~~~~~~~~~~~~~~~~~~~~
\alpha_S(m_Z) \, = \, 0.130 \, \pm \, 0.008
~~~~~~~~~~~~~~~~~~~~~~~~~
\left( E_{\gamma}: \,{\rm BABAR},~\sigma_\gamma \, = \, 100~{\rm MeV}\right),
\eeq
with $\chi^2 \, = \, 8.0$.
Let us note that $\sigma_\gamma$ and $\alpha_S(m_Z)$ are slightly
anti-correlated because by increasing $\alpha_S(m_Z)$ more radiation
is emitted with a smearing effect similar to the one of increasing
$\sigma_\gamma$.

\begin{figure}[th]
\centerline{\epsfxsize=4truein \epsfbox{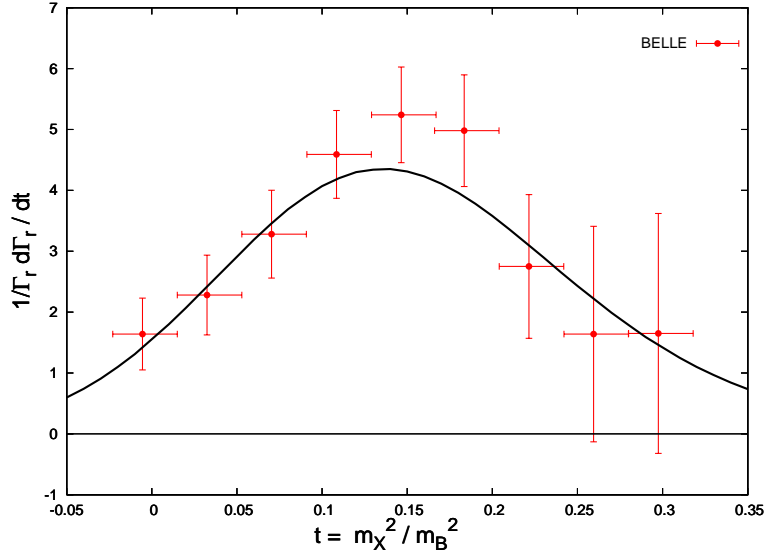}}
\caption{
\small
$B \to X_s \gamma$ photon spectrum from Belle compared to our model for
$\alpha_S(m_Z) \, = \, 0.130$ and $\sigma_\gamma \, = \, 200$ MeV.
}
\label{efbelle}
\end{figure}

The same analysis on the BaBar photon spectrum can be repeated for the Belle
one \cite{bellegam1} (see Fig.~\ref{efbelle}).
The minimization of $H(\sigma_\gamma)$ gives in this case
$\sigma_\gamma \, \simeq \, 200$ MeV.
We obtain a minimum of $\chi^2 \, = \, 5.3$ for $8$
d.o.f., to give
\beq
~~~~~~~~~~~~~
\alpha_S(m_Z) \, = \, 0.130 \, \pm \, 0.005
~~~~~~~~~~~~~~~~~~~~~ \left( E_{\gamma}: \, {\rm BELLE},~\sigma_\gamma \, = \, 200~{\rm MeV} \right).
\eeq

\clearpage

The over-all picture is that there is a good agreement of our model with the data
in the region $m_X  >  1$ GeV, below which single resonances such as $K$ and $K^*$
are expected to have a substantial effect in the dynamics.
The extracted values of $\alpha_S(m_Z)$ are in agreement with the world average
\beq
~~~~~~~~~~~~~~~~
\alpha_S(m_Z) \, = \, 0.1176 \, \pm \, 0.0020
~~~~~~~~~~~~~~~~~~~~~~~~ {\rm \left( PDG 06\right). }
\eeq
at most within two standard deviations.

\section{Semileptonic decay}
\label{section5}

Resummed expressions for the triple-differential distribution
in the inclusive charmless semileptonic $B$ decays,
\beq
\label{sldec}
B \, \to \, X_u \, + \, l \, + \, \nu,
\eeq
as well as for many double and single distributions have been given in
\cite{me,noi1,noi2,noi3}, so we do not repeat them here
\footnote{
A slightly different formalism, which uses light-cone variables
and is equivalent to ours in leading twist, has been developed in
\cite{Lange:2005yw}.}.
To compare with semileptonic data, we just supplement these resummed
expressions with the QCD form factor $\sigma$ computed within our model.

\subsection{Hadron mass distribution}
\begin{figure}[th]
\centerline{\epsfxsize=4truein \epsfbox{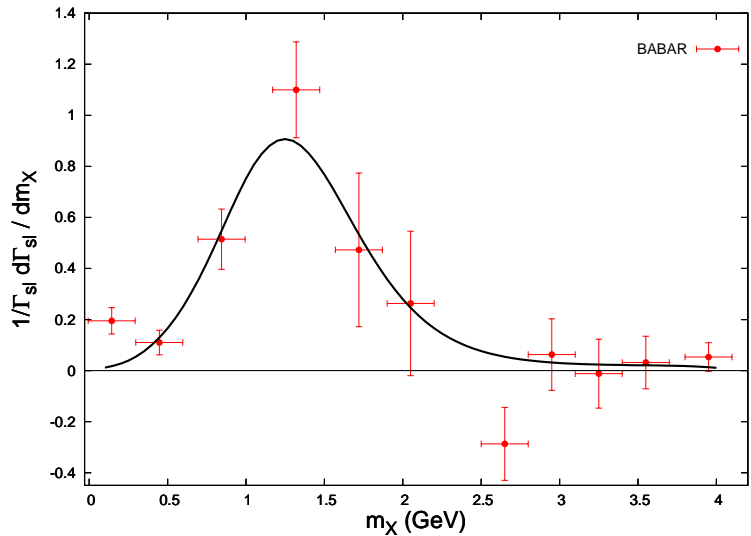}}
\caption{
\small
invariant hadron mass distribution in semileptonic decays from BaBar
compared to our model for $\alpha_S(m_Z) \, = \, 0.119$.
}
\label{slmxbab}
\end{figure}

In Fig.~\ref{slmxbab} we compare the invariant hadron mass distribution
in the semileptonic decay (\ref{sldec}) in our model with data from the
BaBar collaboration \cite{Aubert:2006qi}.
We discard the point with $m_X \, < \, 400$ MeV, which is dominated by the
$\pi$ peak, and the points with $m_X \, > \, 2.6$ GeV, which give basically no
information on the signal
\footnote{
Let us stress that for $m_X \, > \, 1.7$ GeV experimental errors become very large
because of the large background coming from semileptonic $b \, \to \, c$
transitions which have $m_X \, \ge \, m_D \, = \, 1.867$ GeV.
}.
We obtain a minimum $\chi^2 \, = \, 1.1$ for $5$ d.o.f. and, using the
method of the previous section, we obtain:
\beq
~~~~~~~~~~~~~~~
\alpha_S(m_Z) \, = \, 0.119 \, \pm \, 0.003
~~~~~~~~~~~~~~~~~~~~~~~~ \left( m_{Xu}: \, {\rm BABAR } \right).
\eeq
Since the $\rho$ width is larger than that of the $K^*$,
$\Gamma_{\rho} \, \simeq \, 150$ MeV $\simeq \, 3 \Gamma_{K^*}$,
and the binning is rather large ($\Delta m_X \, = \, 400$ MeV),
we do not apply any smearing procedure in this case.
\begin{figure}[th]
\centerline{\epsfxsize=4truein \epsfbox{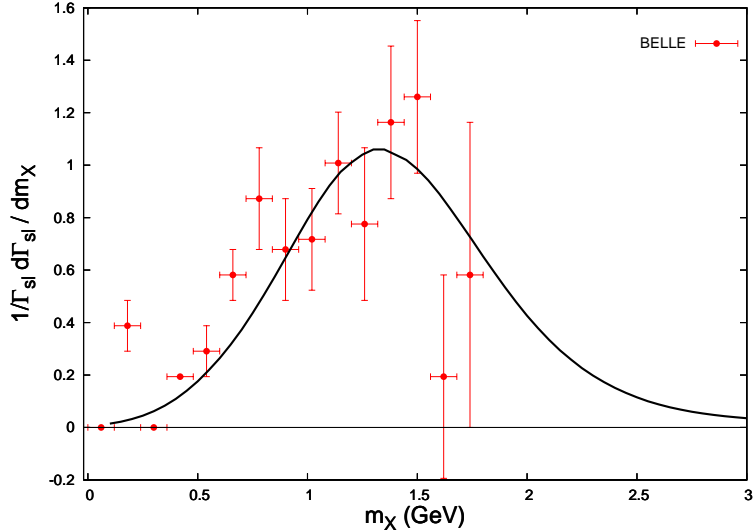}}
\caption{
\small
invariant hadron mass distribution in semileptonic decay from Belle compared to
our model for $\alpha_S(m_Z) \, = \, 0.123$.
}
\label{slmxbel}
\end{figure}

In Fig.~\ref{slmxbel} we present a similar plot with Belle data \cite{bellemxu}.
To extract $\alpha_S(m_Z)$, we discard the first $7$ points, having
$m_X \, < \, 0.8$ GeV. We obtain a minimum
$\chi^2 \, = \, 5.3$ with $7$ d.o.f., to give $\alpha_S(m_Z) \, = \, 0.123 \, \pm \, 0.006$.
Since the binning is smaller for Belle ($\Delta m_X \, = \, 120$ MeV)
than for BaBar, the $\rho$ peak is pretty visible now.
To reduce the resonance effect,
we convolve our theoretical curve {\it and} the experimental data with a normal distribution
of $\sigma \, = \, 300$ MeV,
as we have made with the $m_{Xs}$ spectrum in the previous section.
By discarding the first four points,
we obtain a minimum $\chi^2 \, = \, 0.41$ for 10 d.o.f. to give
$\alpha_S(m_Z) \, = \, 0.115 \, \pm \, 0.004$.
Combining the above measures as we have made for the $m_{Xs}$ spectrum, we quote:
\beq
~~~~~~~~~~~~~~~
\alpha_S(m_Z) \, = \, 0.119 \, \pm \, 0.004
~~~~~~~~~~~~~~~~~~~~~~~~~~~~~~~~~~~~ \left( m_{Xu}: \, {\rm BELLE } \right) .
\eeq
Let us note that semileptonic distributions peak at smaller hadron masses
than radiative ones because they have a smaller average hadron energy \cite{noi1}:
\beq
\langle E_X \rangle_{sl} \, \simeq \, 0.7 \,  \langle E_X \rangle_{rd}.
\eeq
For $\alpha_S(m_Z) \, = \, 0.123$, we find for the peak positions in our model:
\beq
m_{Xu} \, \approx \, 1.3~{\rm GeV}
~~~~~{\rm while}~~~~~
m_{Xs} \, \approx \, 1.7~{\rm GeV}.
\eeq

We end this section by noting that the extracted values of $\alpha_S(m_Z)$
from the above measurements are in agreement with each other as well with the reference
value within one standard deviation.

\subsection{Electron spectrum}

The electron spectrum in the decay (\ref{sldec}) is affected by a large background for
\beq
E_e \, < \,  \frac{m_B}{2} \, \left( 1 \, - \, \frac{m_D^2}{m_B^2} \right)
\, \simeq \, 2.31 \, {\rm GeV}
\eeq
coming from the decays
\beq
B \, \to \, X_c \, + \, l \, + \, \nu.
\eeq
This background is larger than the signal by two orders of magnitude
because $|V_{ub}|^2/|V_{cb}|^2 \sim 10^{-2}$.
In order to avoid the large errors coming from its subtraction,
we have normalized the theory and the data to one in the region
$E_e \, > \, 2.31$ GeV.
Instead of the electron energy, we prefer to use the variable
\beq
\bar{x}_e \, \equiv \, 1 \, - \, \frac{2 E_e}{m_B},
\eeq
which is equal to zero for the largest electron energy. The charm
background occurs for $\bar{x}_e \, > \, 0.125$.
To include the Doppler effect, we convolve our spectra
with a normal distribution of standard deviation $\sigma_e \, = \, 100$ MeV.

\begin{figure}[th]
\centerline{\epsfxsize=4truein \epsfbox{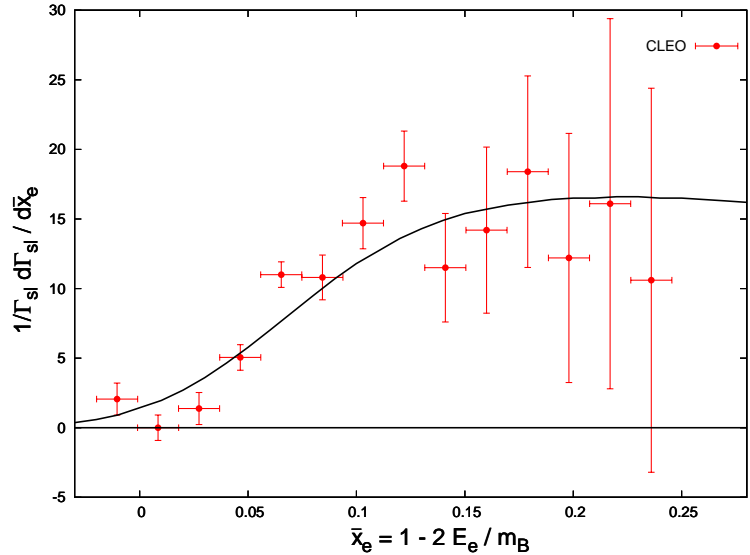}}
\caption{
\small
electron spectrum in semileptonic charmless $B$ decay from Cleo compared to our model
with $\alpha_S(m_Z) \, = \, 0.118$.
The data and the theory are normalized to one in the charm background free region
$0 \, < \, \bar{x}_e \, < \, 0.125$.
}
\label{eecleo}
\end{figure}

In Fig.~\ref{eecleo} we compare the electron spectrum in our model
\footnote{ Let us note that the tree-level electron spectrum has a
maximum for $\bar{x}_e = 0$, at the largest electron energy, where
it is flat. The shift of the maximum inside the kinematical
domain, in $\bar{x}_e \approx 0.2$, is a Sudakov effect. Because
of infrared divergencies, soft radiation is always emitted and the
high energy electron recoils against a neutrino and a massive
$up$-quark jet, instead of a massless one. } with data from the
Cleo Collaboration \cite{cleoel}. We obtain a minimum $\chi^2 \, =
\, 30$ for $13$ d.o.f. and with the analysis described above we
obtain: \beq ~~~~~~~~~~~~~~~~ \alpha_S(m_Z) \, = \, 0.117 \, \pm
\, 0.005 ~~~~~~~~~~~~~~~~~~~~~ \left( E_e: \, {\rm CLEO} \right).
\eeq The over-all agreement of the model with the data is
acceptable in all the measured range of electron energies. In the
region affected by the charm background, experimental errors
become however very large.

\begin{figure}[th]
\centerline{\epsfxsize=4truein \epsfbox{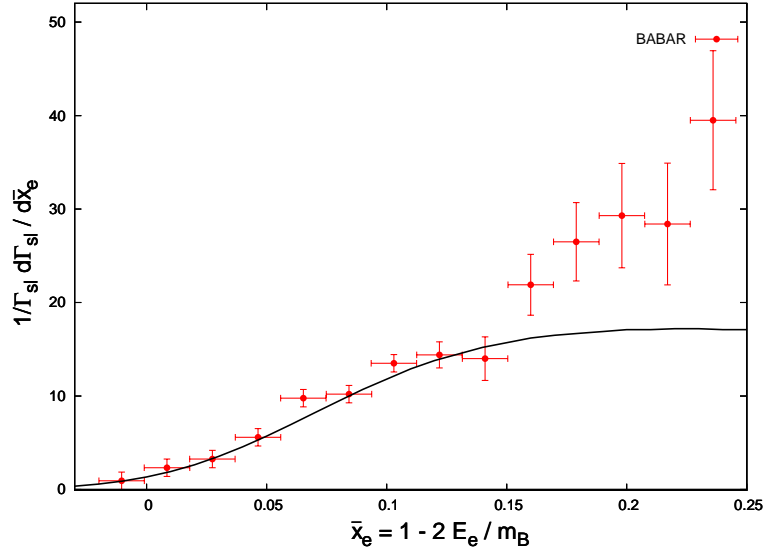}}
\caption{
\small
electron spectrum in semileptonic decay from BaBar compared to our model
with $\alpha_S(m_Z) \, = \, 0.119$.
The data and the theory are normalized to one in the charm background free region
$0 \, < \, \bar{x}_e \, < \, 0.125$.
}
\label{eebabar}
\end{figure}

In Fig.~\ref{eebabar} we compare our prediction with the electron spectrum
measured by the BaBar collaboration \cite{babarel}.
In the $\chi^2$ analysis we remove the $4$ points with the smallest electron
energies, which are affected by the subtraction of the charm background.
We obtain a minimum $\chi^2 \, = \, 16$ for $9$ d.o.f. and we obtain
\beq
~~~~~~~~~~~~~~~~
\alpha_S(m_Z) \, = \, 0.119 \, \pm \, 0.005
~~~~~~~~~~~~~~~~~ \left( E_e: \, {\rm BABAR} \right).
\vspace{-0.5cm}
\eeq

\begin{figure}[th]
\centerline{\epsfxsize=4truein \epsfbox{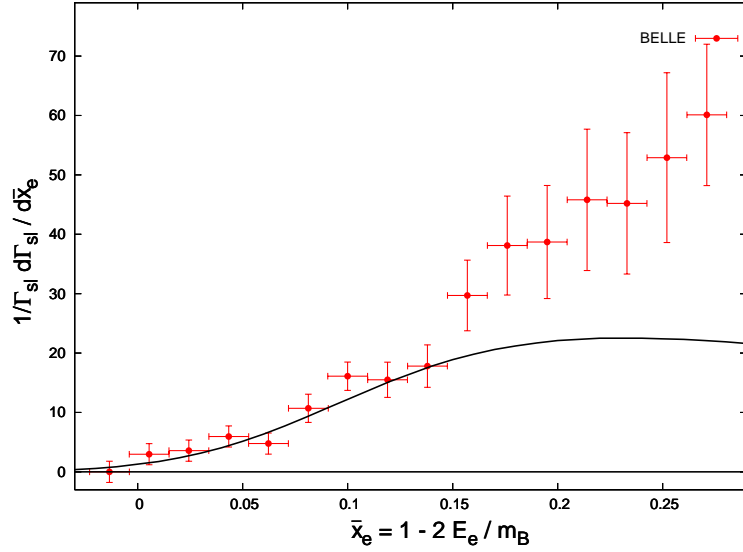}}
\caption{
\small
electron spectrum in semileptonic decay from Belle compared to our model
for $\alpha_S(m_Z) \, = \, 0.135$.
The data and the theory are normalized to one in the charm background free region
$0 \, < \, \bar{x}_e \, < \, 0.125$.
}
\label{eebelle}
\end{figure}

In Fig.~\ref{eebelle} we compare our model with Belle data \cite{belleel}.
For the $\chi^2$ analysis we have discarded the seven points with the largest
$\bar{x}_e$, i.e. with smallest electron energies.
We obtain a minimum $\chi^2 \, = \, 7$ for $8$ d.o.f. for
$\alpha_S(m_Z) \, \approx \, 0.135$.
Since the $\chi^2$ is a rather irregular function of $\alpha_S(m_Z)$ in this
case --- without a well-shaped minimum --- we are not able to estimate
the error.

\begin{figure}[th]
\centerline{\epsfxsize=4truein \epsfbox{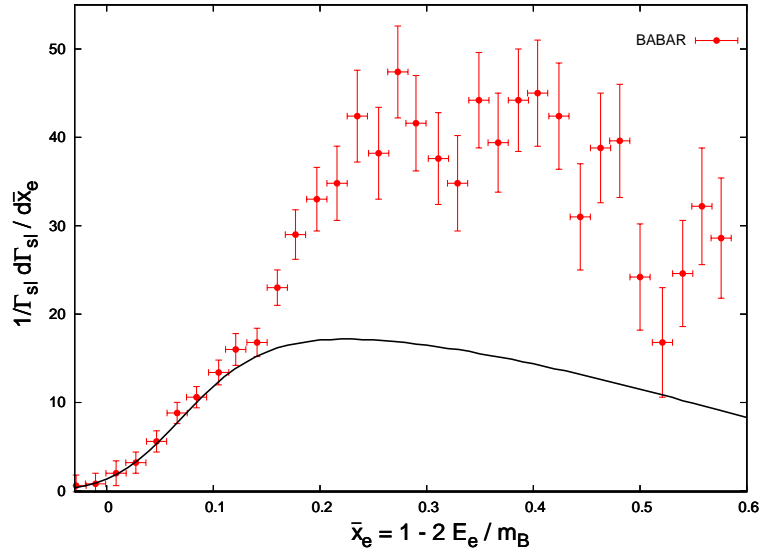}}
\caption{
\small
electron spectrum in semileptonic decay extending down to $1.1$ GeV from BaBar
(preliminary) compared to our model with $\alpha_S(m_Z) \, = \, 0.119$.
The data and the theory are normalized to one in the charm background free region
$0 \, < \, \bar{x}_e \, < \, 0.125$.
}
\label{eecompl}
\end{figure}

Finally, in Fig.~\ref{eecompl} we compare our model for $\alpha_S(m_Z)\, = \, 0.119$ with a preliminary
measure of the electron spectrum of the BaBar collaboration extending down to $E_e \, = \, 1.1$ GeV
\cite{Aubert:2004bv}.
As it is clearly seen, the theoretical spectrum is harder than the experimental
one. We do not known whether this discrepancy is related to a deficiency of our
model or to an under-estimated charm background.
\clearpage

We may summarize our analysis of the electron spectra by saying that
the agreement theory-data is less clear in this case.
The agreement is acceptable in the charm background free region,
i.e. for $2.31 \, < \, E_e \, < \, 2.64$ GeV; the errors in the measure of
$\alpha_S(m_Z)$ are however larger than in previous cases.
There is not a good agreement with the preliminary BaBar spectrum for small
electron energies: our model predicts a broad maximum around $E_e \, = \, 2.1$ GeV,
while the data seem to peak at lower energies.

\section{Conclusions}
\label{section6}

We have presented a model for the QCD form factor describing
radiative and semileptonic $B$ decay spectra based on
soft-gluon resummation to next-to-next-to-leading logarithmic accuracy
and on a power expansion in an ghost-less time-like coupling.
The latter is free from infrared singularities (Landau ghost)
and resumes absorptive effects in gluon cascades to all orders.

The agreement with invariant hadron mass distributions in radiative
and semi-leptonic decays measured by Cleo, BaBar and Belle is in
general a good one. The $\chi^2$/d.o.f. values are acceptable and the
extracted values of $\alpha_S(m_Z)$ are in agreement with the current
PDG average within two at most standard deviations.

The agreement with the electron spectra in semi-leptonic decays is
in general is, in general, not as good. Even restricting the
analysis to the end-point region free from the charm background
($2.31 \div 2.64$ GeV),  $\chi^2$/d.o.f. values are larger and the
extracted values of $\alpha_S(m_Z)$ are generally less accurate
than in previous cases. The preliminary BaBar measure of the
electron spectrum down to $1.1$ GeV is not in good agreement with
our model, which predicts a harder spectrum, with a broad maximum
around $2.1$ GeV. We do not know whether the discrepancy is to be
attributed to a deficiency of our model or to an under-subtracted
charm background. In the former case, one could think to a
non-perturbative component which is accidentally larger in the
electron spectrum than in other semi-leptonic or radiative
spectra.

In general, the model seems to work quite well, validating the
idea that Fermi-motion effects can be described in a resummed pQCD
framework with an effective QCD coupling, which remains reasonably
smaller than one in the relevant integration domain.
Since the effective coupling is constructed by means of an extrapolation
at low energy of the standard coupling, non-perturbative Fermi-motion effects are
connected in a smooth way to the perturbative ones --- namely soft gluon radiation ---
in our model.
More accurate data on any distribution sensitive to soft-gluon effects
are needed to put the model to a stringent test.
Theoretical predictions could be sharpened in the future by including
second-order corrections to the coefficient functions
and remainder functions, as soon as they become available;
that would allow to work within a complete NNLO approximation.

We have found that the inclusion of the NNLO effects in our model
is crucial for a good description of the data; the model could be
improved by including NNNLO terms. We have also found that the
non-power expansion proposed in \cite{shirkov,Aglietti:2004fz}
does not accurately describe soft-gluon effects.

Let us end with a general comment. We find it remarkable that with
such a simple model as the one we have formulated, without any
adjustable parameter, it is possible to extract reasonable values
of $\alpha_S(m_Z)$ from spectra with a hard scale of just a few
GeV.

\vskip 1.truecm
\centerline{\bf Acknowledgments}

\vskip .2truecm

We would like to thank G.~D'Agostini, R.~Faccini and
M.~Pierini for discussions.
\vskip 1.truecm
\vspace{1cm}
\appendix

\section{Coefficient Functions and Remainder Functions for the Radiative Decay}
\allowdisplaybreaks

The coefficient functions $C_i$ of the local operators $O_i$
entering the effective Hamiltonian
\beq
{\cal H}^{b\to s \gamma} \, = \, \sum_{i=1}^8 C_i \, O_i
\eeq
can be taken as:
\begin{equation}
(C_1, \, C_2, \, C_3, \, C_4, \, C_5, \, C_6, \, C_7, \, C_8) \, = \,
(- \, 0.480,\, 1.023,\, - \, 0.0045,\, - \, 0.0640,\, 0.0004,\, 0.0009,\, - \, 0.304,\, - \, 0.148).
\end{equation}
The analytic expressions of the $C_i$'s
as functions of $m_b, \, m_t, \, m_W$ and $\alpha_S(m_Z)$ as well as of the
$O_i$'s can be found in \cite{misiak}
\footnote{
Unlike previous works (\cite{me} and \cite{acg}), we always insert in the formulas the corrected
$C_7\, = \, C_7^{(0)} \, + \, \alpha_S/(4\pi) \, C_7^{(1)}$.
}.
The coefficients $C_i$ for $i = 3,4,5,6$ are very small, implying that the contributions
of the related operators can be neglected.

The functions entering the leading-order remainder function $D_r^{(1)}(t)$
read in our conventions:
\begin{eqnarray}
f_{11}(t) &=& + \, \frac{1}{36} \, f_{22}(t) ;
\\
f_{12}(t) &=& - \, \frac{1}{3} \, f_{22}(t) ;
\\
f_{17}(t) &=& - \, \frac{1}{6} \, f_{27}(t) ;
\\
f_{18}(t) &=& - \, \frac{1}{6} \, f_{28}(t) ;
\\
f_{22}(t) &=&
+ \, \frac{16}{27} \,  k \,
\left\{
t \int_0^{(1-t)/k} dv \, (1-k \, v) \,
\left| \frac{G(v)}{v} + \frac{1}{2} \right|^2
+ \int_{(1-t)/k}^{1/k} dv \, (1-k \, v)^2
\left| \frac{G(v)}{v} + \frac{1}{2} \right|^2
\right\} ;
\\
f_{27}(t) &=&
- \, \frac{8}{9} \, k^2 \,
\left\{
t \int_0^{(1-t)/k} dv \, {\rm Re} \, \left[ G(v) + \frac{v}{2} \right]
+ \int_{(1-t)/k}^{1/k} dv \, (1-k \, v) \, {\rm Re}\left[ G(v) + \frac{v}{2} \right]
\right\} ;
\\
f_{28}(t) &=& - \, \frac{1}{3} \, f_{27}(t) ;
\\
f_{77}(t) &=&
+ \, \frac{t}{9} \,\left[ 30 + 3\,t - 2\,t^2 - 3\,\left( 4 - t \right) \,\log t \right];
\\
f_{78}(t) &=&
- \, \frac{2}{27} \,
\left[ 2\,{\pi }^2 - 27\,t + 3\,t^2 -
      t^3 + 12\,t\,\log t - 12 \, {\rm Li}_2(1 - t)
\right];
\\
f_{88}(t) &=&
 +\, \frac{1}{81}
\Big\{
\,- \, 2 {\pi }^2 +
    t\,\left[ 21 +
       \left( 9 - 2\,t \right) \,t \right]  +
    24\,\log (1 - t) -
    6\,\log \frac{m_b}{m_s}\,
     \left[ t\,
        \left( 2 + t \right)  + 4\,\log (1 - t)
       \right] \, +
\nonumber\\
&& ~~~~~~~ - \, 3\,t\,\left( 2 + t \right) \,
     \log t + 12 \, {\rm Li}_2(1 - t)
\Big\}.
\end{eqnarray}
We have defined:
\begin{equation}
G(v) \, = \, \Bigg\{
\begin{array}{ll}
- 2 \, {\rm arctan}^2 \left(\sqrt{\frac{v}{4-v}}   \right) &  v < 4 ;
\\
2 \, \log^2 \left[\frac{ \sqrt{v} + \sqrt{v-4} }{2}\right]
- 2 \pi i \log \left[\frac{\sqrt{v}+\sqrt{v-4}}{2}\right]
-\frac{\pi^2}{2}
&  v \ge 4,
\end{array}
\end{equation}
and
\begin{equation}
k \, = \, \frac{m_c^2}{m_b^2}.
\end{equation}

\vspace{1cm}
\section{QCD form factor}

In this appendix we tabulate the values of the QCD form factor $\sigma(u;\,w)$
in our model as a function of the infrared variable
\beq
u \, \equiv \, \frac{E_X - \sqrt{E_X^2 - m_X^2}}{ E_X + \sqrt{E_X^2 - m_X^2} }
\, \simeq \, \frac{m_X^2}{4 E_X^2} ~~~~~~~~~\left( {\rm for}~ m_X \, \ll \, E_X \right)
\eeq
and of the total final hadron energy
\beq
w \, \equiv \, \frac{2 E_X}{m_B}
\eeq
for $\alpha_S(m_Z) \, = \, 0.115, \, 0.120$ and  $0.125$.
The hard scale in the process is
\beq
Q \, = \, w \, m_B.
\eeq
In the radiative case one sets $t = u$ and $w = 1$ while in the semileptonic case
the form factor as a function of $w$ ($0 < w < 1$) is needed \cite{noi1,noi2,noi3}.
By using the following tables, the reader can obtain the form
factor for all the values of $u$ and $w$ by means of a straightforward interpolation,
avoiding the delicate numerical integrations related to the Mellin transform and to
the inverse Mellin transform.
In agreement with physical intuition, by lowering the hard scale,
the peak of the form factor broadens and shifts to larger $u$'s because of the
coupling growth.

\clearpage
\begin{center}
{$\bf{ \sigma(u; \, w): ~ \alpha_S(m_Z) \, = \, 0.115}$ } \\*
\end{center}
\renewcommand{\tabcolsep}{0.26cm}
\begin{tabular}{|c||c|c|c|c|c|c|}
\hline \hline
$u$ &  $w=0.10$ & $w=0.28$ & $w=0.46$ & $w=0.64$ & $w=0.82$ & $w=1.00$ \\
\hline \hline
$0.01$&$2.89\times 10^{\text{-6}}$&$5.094\times 10^{\text{-6}}$&$1.218\times 10^{\text{-5}}$&$1.088\times 10^{\text{-4}}$&$8.073\times 10^{\text{-4}}$&$3.825\times 10^{\text{-3}}$\\ \hline
$0.02$&$3.063\times 10^{\text{-6}}$&$1.028\times 10^{\text{-4}}$&$3.525\times 10^{\text{-3}}$&$3.077\times 10^{\text{-2}}$&$1.344\times 10^{\text{-1}}$&$3.923\times 10^{\text{-1}}$\\ \hline
$0.03$&$3.686\times 10^{\text{-6}}$&$3.574\times 10^{\text{-3}}$&$6.732\times 10^{\text{-2}}$&$3.688\times 10^{\text{-1}}$&$1.12\
$&$2.426$\\ \hline
$0.04$&$4.715\times 10^{\text{-5}}$&$3.068\times 10^{\text{-2}}$&$3.597\times 10^{\text{-1}}$&$1.407\
$&$3.295$&$\
5.788$\\ \hline
$0.05$&$4.848\times 10^{\text{-4}}$&$1.277\times 10^{\text{-1}}$&$1.03\
$&$3.102$&$5.948\
$&$8.903$\\ \hline
$0.06$&$2.577\times 10^{\text{-3}}$&$3.485\times 10^{\text{-1}}$&$2.065\
$&$5.029$&$\
8.209$&$1.082\times 10^1$\\ \hline
$0.07$&$9.099\times 10^{\text{-3}}$&$7.256\times 10^{\text{-1}}$&$3.313\
$&$6.749$&$\
9.64$&$1.145\times 10^1$\\ \hline
$0.08$&$2.451\times 10^{\text{-2}}$&$1.256\
$&$4.578$&$\
8.002$&$1.021\times 10^1$&$1.111\times 10^1$\\ \hline
$0.09$&$5.449\times 10^{\text{-2}}$&$1.905\
$&$5.697$&$\
8.713$&$1.008\times 10^1$&$1.019\times 10^1$\\ \hline
$0.10$&$1.052\times 10^{\text{-1}}$&$2.621\
$&$6.568$&$\
8.929$&$9.493\
$&$9.004$\\ \hline
$0.11$&$1.821\times 10^{\text{-1}}$&$3.346\
$&$7.152$&$\
8.754$&$8.637\
$&$7.748$\\ \hline
$0.12$&$2.894\times 10^{\text{-1}}$&$4.031\
$&$7.456$&$\
8.304$&$7.661\
$&$6.549$\\ \hline
$0.13$&$4.296\times 10^{\text{-1}}$&$4.636\
$&$7.511$&$\
7.681$&$6.669\
$&$5.469$\\ \hline
$0.14$&$6.029\times 10^{\text{-1}}$&$5.135\
$&$7.366$&$\
6.964$&$5.725\
$&$4.531$\\ \hline
$0.15$&$8.075\times 10^{\text{-1}}$&$5.516\
$&$7.067$&$\
6.213$&$4.863\
$&$3.734$\\ \hline
$0.16$&$1.04\
$&$5.776$&$\
6.661$&$5.472\
$&$4.099$&$\
3.065$\\ \hline
$0.17$&$1.295\
$&$5.92$&$\
6.186$&$4.769$&$\
3.435$&$2.51\
$\\ \hline
$0.18$&$1.568\
$&$5.958$&$\
5.673$&$4.119\
$&$2.864$&$\
2.05$\\ \hline
$0.19$&$1.851\
$&$5.902$&$\
5.145$&$3.532\
$&$2.379$&$\
1.67$\\ \hline
$0.20$&$2.138\
$&$5.769$&$\
4.621$&$3.008\
$&$1.968$&$\
1.356$\\ \hline
$0.21$&$2.423\
$&$5.571$&$\
4.114$&$2.548\
$&$1.621$&$\
1.097$\\ \hline
$0.22$&$2.7\
$&$5.325$&$3.635\
$&$2.146$&$\
1.329$&$8.814\times 10^{\text{-1}}$\\ \hline
$0.23$&$2.963\
$&$5.042$&$\
3.187$&$1.798\
$&$1.083$&$\
7.027\times 10^{\text{-1}}$\\ \hline
$0.24$&$3.208\
$&$4.734$&$\
2.776$&$1.498\
$&$8.765\times 10^{\text{-1}}$&$5.537\times 10^{\text{-1}}$\\ \hline
$0.25$&$3.43\
$&$4.411$&$\
2.401$&$1.24\
$&$7.021\times 10^{\text{-1}}$&$4.293\times 10^{\text{-1}}$\\ \hline
$0.26$&$3.627\
$&$4.081$&$\
2.063$&$1.019\
$&$5.55\times 10^{\text{-1}}$&$3.249\times 10^{\text{-1}}$\\ \hline
$0.27$&$3.797\
$&$3.75$&$\
1.761$&$8.29\times 10^{\text{-1}}$&$4.307\times 10^{\text{-1}}$&$2.372\times 10^{\text{-1}}$\\ \hline
$0.28$&$3.937\
$&$3.423$&$\
1.491$&$6.665\times 10^{\text{-1}}$&$3.255\times 10^{\text{-1}}$&$1.632\times 10^{\text{-1}}$\\ \hline
$0.29$&$4.048\
$&$3.105$&$\
1.253$&$5.273\times 10^{\text{-1}}$&$2.363\times 10^{\text{-1}}$&$1.005\times 10^{\text{-1}}$\\ \hline
$0.30$&$4.128\
$&$2.798$&$\
1.042$&$4.079\times 10^{\text{-1}}$&$1.606\times 10^{\text{-1}}$&$4.656\times 10^{\text{-2}}$\\ \hline
$0.31$&$4.18\
$&$2.504$&$\
8.571\times 10^{\text{-1}}$&$3.057\times 10^{\text{-1}}$&$9.62\times 10^{\text{-2}}$&$1.027\times 10^{\text{-4}}$\\ \hline
$0.32$&$4.202\
$&$2.227$&$\
6.948\times 10^{\text{-1}}$&$2.18\times 10^{\text{-1}}$&$4.139\times 10^{\text{-2}}$&$-3.977\times 10^{\text{-2}}$\\ \hline
$0.33$&$4.197\
$&$1.965$&$\
5.528\times 10^{\text{-1}}$&$1.427\times 10^{\text{-1}}$&$-5.313\times 10^{\text{-3}}$&$-7.377\times 10^{\text{-2}}$\\ \hline
$0.34$&$4.166\
$&$1.722$&$\
4.29\times 10^{\text{-1}}$&$7.814\times 10^{\text{-2}}$&$-4.524\times 10^{\text{-2}}$&$-1.026\times 10^{\text{-1}}$\\ \hline
$0.35$&$4.112\
$&$1.496$&$\
3.211\times 10^{\text{-1}}$&$2.272\times 10^{\text{-2}}$&$-7.955\times 10^{\text{-2}}$&$-1.27\times 10^{\text{-1}}$\\ \hline
$0.36$&$4.035\
$&$1.287$&$\
2.274\times 10^{\text{-1}}$&$-2.485\times 10^{\text{-2}}$&$-1.093\times 10^{\text{-1}}$&$-1.476\times 10^{\text{-1}}$\\ \hline
$0.37$&$3.938\
$&$1.096$&$\
1.463\times 10^{\text{-1}}$&$-6.568\times 10^{\text{-2}}$&$-1.353\times 10^{\text{-1}}$&$-1.649\times 10^{\text{-1}}$\\ \hline
$0.38$&$3.824\
$&$9.216\times 10^{\text{-1}}$&$7.613\times 10^{\text{-2}}$&$-1.007\times 10^{\text{-1}}$&$-1.578\times 10^{\text{-1}}$&$-1.796\times 10^{\text{-1}}$\\ \hline
$0.39$&$3.693\
$&$7.63\times 10^{\text{-1}}$&$1.552\times 10^{\text{-2}}$&$-1.307\times 10^{\text{-1}}$&$-1.771\times 10^{\text{-1}}$&$-1.918\times 10^{\text{-1}}$\\ \hline
$0.40$&$3.55\
$&$6.195\times 10^{\text{-1}}$&$-3.678\times 10^{\text{-2}}$&$-1.564\times 10^{\text{-1}}$&$-1.936\times 10^{\text{-1}}$&$-2.021\times 10^{\text{-1}}$\\ \hline
$0.50$&$1.791\
$&$-1.934\times 10^{\text{-1}}$&$-2.85\times 10^{\text{-1}}$&$-2.775\times 10^{\text{-1}}$&$-2.596\times 10^{\text{-1}}$&$-2.42\times 10^{\text{-1}}$\\ \hline
$0.60$&$3.035\times 10^{\text{-1}}$&$-3.989\times 10^{\text{-1}}$&$-3.229\times 10^{\text{-1}}$&$-2.851\times 10^{\text{-1}}$&$-2.551\times 10^{\text{-1}}$&$-2.346\times 10^{\text{-1}}$\\ \hline
$0.70$&$-4.713\times 10^{\text{-1}}$&$-4.072\times 10^{\text{-1}}$&$-3.088\times 10^{\text{-1}}$&$-2.615\times 10^{\text{-1}}$&$-2.335\times 10^{\text{-1}}$&$-2.15\times 10^{\text{-1}}$\\ \hline
$0.80$&$-7.243\times 10^{\text{-1}}$&$-3.596\times 10^{\text{-1}}$&$-2.718\times 10^{\text{-1}}$&$-2.311\times 10^{\text{-1}}$&$-2.076\times 10^{\text{-1}}$&$-1.92\times 10^{\text{-1}}$\\ \hline
$0.90$&$-6.43\times 10^{\text{-1}}$&$-2.953\times 10^{\text{-1}}$&$-2.286\times 10^{\text{-1}}$&$-1.975\times 10^{\text{-1}}$&$-1.792\times 10^{\text{-1}}$&$-1.653\times 10^{\text{-1}}$\\ \hline
$0.98$&$-4.249\times 10^{\text{-1}} $&$ - 2.458\times 10^{\text{-1}} $&$ -
              2.017\times 10^{\text{-1}} $&$ - 1.803\times 10^{\text{-1}}
$&$ -
      1.672\times 10^{\text{-1}} $&$ - 1.58\times 10^{\text{-1}}$\\
 \hline
\end{tabular}

\vglue1.5cm
\begin{center}
{$\bf{ \sigma(u; \, w): ~ \alpha_S(m_Z) \, = \, 0.120}$ } \\*
\end{center}
\renewcommand{\tabcolsep}{0.26cm}
\begin{tabular}{|c||c|c|c|c|c|c|}
\hline \hline
$u$ &  $w=0.10$ & $w=0.28$ & $w=0.46$ & $w=0.64$ & $w=0.82$ & $w=1.00$ \\
\hline \hline
$0.01$&$2.575\times 10^{\text{-6}}$&$4.491\times 10^{\text{-6}}$&$6.691\times 10^{\text{-6}}$&$2.153\times 10^{\text{-5}}$&$1.321\times 10^{\text{-4}}$&$6.542\times 10^{\text{-4}}$\\ \hline
$0.02$&$2.748\times 10^{\text{-6}}$&$2.286\times 10^{\text{-5}}$&$7.371\times 10^{\text{-4}}$&$7.231\times 10^{\text{-3}}$&$3.539\times 10^{\text{-2}}$&$1.147\times 10^{\text{-1}}$\\ \hline
$0.03$&$2.992\times 10^{\text{-6}}$&$8.639\times 10^{\text{-4}}$&$1.881\times 10^{\text{-2}}$&$1.185\times 10^{\text{-1}}$&$4.081\times 10^{\text{-1}}$&$9.901\times 10^{\text{-1}}$\\ \hline
$0.04$&$1.341\times 10^{\text{-5}}$&$8.992\times 10^{\text{-3}}$&$1.247\times 10^{\text{-1}}$&$5.674\times 10^{\text{-1}}$&$1.517\
$&$2.995$\\ \hline
$0.05$&$1.359\times 10^{\text{-4}}$&$4.372\times 10^{\text{-2}}$&$4.23\times 10^{\text{-1}}$&$1.493\
$&$3.282$&$\
5.527$\\ \hline
$0.06$&$8.021\times 10^{\text{-4}}$&$1.357\times 10^{\text{-1}}$&$9.753\times 10^{\text{-1}}$&$2.797\
$&$5.245$&$\
7.775$\\ \hline
$0.07$&$3.121\times 10^{\text{-3}}$&$3.155\times 10^{\text{-1}}$&$1.761\
$&$4.239$&$\
6.961$&$9.285\
$\\ \hline
$0.08$&$9.146\times 10^{\text{-3}}$&$6.013\times 10^{\text{-1}}$&$2.695\
$&$5.582$&$\
8.182$&$9.98\
$\\ \hline
$0.09$&$2.192\times 10^{\text{-2}}$&$9.932\times 10^{\text{-1}}$&$3.67\
$&$6.663$&$8.849\
$&$9.991$\\ \hline
$0.10$&$4.521\times 10^{\text{-2}}$&$1.475\
$&$4.587$&$\
7.407$&$9.021\
$&$9.516$\\ \hline
$0.11$&$8.315\times 10^{\text{-2}}$&$2.018\
$&$5.373$&$\
7.812$&$8.805\
$&$8.74$\\ \hline
$0.12$&$1.397\times 10^{\text{-1}}$&$2.589\
$&$5.986$&$\
7.916$&$8.316\
$&$7.817$\\ \hline
$0.13$&$2.181\times 10^{\text{-1}}$&$3.157\
$&$6.41$&$\
7.774$&$7.657$&$\
6.855$\\ \hline
$0.14$&$3.208\times 10^{\text{-1}}$&$3.692\
$&$6.65$&$\
7.447$&$6.911$&$\
5.924$\\ \hline
$0.15$&$4.489\times 10^{\text{-1}}$&$4.171\
$&$6.724$&$\
6.991$&$6.141\
$&$5.062$\\ \hline
$0.16$&$6.024\times 10^{\text{-1}}$&$4.58\
$&$6.657$&$6.453\
$&$5.388$&$\
4.29$\\ \hline
$0.17$&$7.799\times 10^{\text{-1}}$&$4.909\
$&$6.474$&$\
5.873$&$4.681\
$&$3.612$\\ \hline
$0.18$&$9.791\times 10^{\text{-1}}$&$5.155\
$&$6.201$&$\
5.282$&$4.033\
$&$3.024$\\ \hline
$0.19$&$1.197\
$&$5.317$&$\
5.862$&$4.702\
$&$3.451$&$\
2.521$\\ \hline
$0.20$&$1.429\
$&$5.401$&$\
5.478$&$4.149\
$&$2.935$&$\
2.093$\\ \hline
$0.21$&$1.671\
$&$5.413$&$\
5.067$&$3.633\
$&$2.482$&$\
1.73$\\ \hline
$0.22$&$1.918\
$&$5.361$&$\
4.643$&$3.159\
$&$2.089$&$\
1.423$\\ \hline
$0.23$&$2.167\
$&$5.254$&$\
4.219$&$2.729\
$&$1.748$&$\
1.163$\\ \hline
$0.24$&$2.412\
$&$5.1$&$\
3.803$&$2.343$&$\
1.454$&$9.444\times 10^{\text{-1}}$\\ \hline
$0.25$&$2.649\
$&$4.909$&$\
3.404$&$1.999\
$&$1.201$&$\
7.596\times 10^{\text{-1}}$\\ \hline
$0.26$&$2.875\
$&$4.689$&$\
3.025$&$1.695\
$&$9.845\times 10^{\text{-1}}$&$6.034\times 10^{\text{-1}}$\\ \hline
$0.27$&$3.085\
$&$4.446$&$\
2.67$&$1.427\
$&$7.986\times 10^{\text{-1}}$&$4.712\times 10^{\text{-1}}$\\ \hline
$0.28$&$3.278\
$&$4.186$&$\
2.34$&$1.193\
$&$6.394\times 10^{\text{-1}}$&$3.593\times 10^{\text{-1}}$\\ \hline
$0.29$&$3.451\
$&$3.916$&$\
2.037$&$9.873\times 10^{\text{-1}}$&$5.031\times 10^{\text{-1}}$&$2.641\times 10^{\text{-1}}$\\ \hline
$0.30$&$3.602\
$&$3.64$&$\
1.761$&$8.082\times 10^{\text{-1}}$&$3.864\times 10^{\text{-1}}$&$1.828\times 10^{\text{-1}}$\\ \hline
$0.31$&$3.73\
$&$3.362$&$\
1.51$&$6.522\times 10^{\text{-1}}$&$2.864\times 10^{\text{-1}}$&$1.127\times 10^{\text{-1}}$\\ \hline
$0.32$&$3.834\
$&$3.086$&$\
1.283$&$5.165\times 10^{\text{-1}}$&$2.008\times 10^{\text{-1}}$&$5.236\times 10^{\text{-2}}$\\ \hline
$0.33$&$3.913\
$&$2.815$&$\
1.08$&$3.986\times 10^{\text{-1}}$&$1.274\times 10^{\text{-1}}$&$5.426\times 10^{\text{-4}}$\\ \hline
$0.34$&$3.968\
$&$2.552$&$\
8.978\times 10^{\text{-1}}$&$2.963\times 10^{\text{-1}}$&$6.449\times 10^{\text{-2}}$&$-4.377\times 10^{\text{-2}}$\\ \hline
$0.35$&$3.999\
$&$2.298$&$\
7.355\times 10^{\text{-1}}$&$2.076\times 10^{\text{-1}}$&$1.037\times 10^{\text{-2}}$&$-8.151\times 10^{\text{-2}}$\\ \hline
$0.36$&$4.007\
$&$2.055$&$\
5.915\times 10^{\text{-1}}$&$1.308\times 10^{\text{-1}}$&$-3.641\times 10^{\text{-2}}$&$-1.136\times 10^{\text{-1}}$\\ \hline
$0.37$&$3.992\
$&$1.824$&$\
4.64\times 10^{\text{-1}}$&$6.426\times 10^{\text{-2}}$&$-7.71\times 10^{\text{-2}}$&$-1.407\times 10^{\text{-1}}$\\ \hline
$0.38$&$3.956\
$&$1.607$&$\
3.515\times 10^{\text{-1}}$&$6.723\times 10^{\text{-3}}$&$-1.125\times 10^{\text{-1}}$&$-1.636\times 10^{\text{-1}}$\\ \hline
$0.39$&$3.9\
$&$1.403$&$\
2.527\times 10^{\text{-1}}$&$-4.299\times 10^{\text{-2}}$&$-1.433\times 10^{\text{-1}}$&$-1.83\times 10^{\text{-1}}$\\ \hline
$0.40$&$3.826\
$&$1.212$&$\
1.659\times 10^{\text{-1}}$&$-8.59\times 10^{\text{-2}}$&$-1.697\times 10^{\text{-1}}$&$-1.993\times 10^{\text{-1}}$\\ \hline
$0.50$&$2.42\
$&$-9.317\times 10^{\text{-3}}$&$-2.727\times 10^{\text{-1}}$&$-2.938\times 10^{\text{-1}}$&$-2.836\times 10^{\text{-1}}$&$-2.655\times 10^{\text{-1}}$\\ \hline
$0.60$&$7.765\times 10^{\text{-1}}$&$-4.072\times 10^{\text{-1}}$&$-3.595\times 10^{\text{-1}}$&$-3.213\times 10^{\text{-1}}$&$-2.858\times 10^{\text{-1}}$&$-2.61\times 10^{\text{-1}}$\\ \hline
$0.70$&$-3.087\times 10^{\text{-1}}$&$-4.693\times 10^{\text{-1}}$&$-3.545\times 10^{\text{-1}}$&$-2.965\times 10^{\text{-1}}$&$-2.617\times 10^{\text{-1}}$&$-2.388\times 10^{\text{-1}}$\\ \hline
$0.80$&$-7.857\times 10^{\text{-1}}$&$-4.228\times 10^{\text{-1}}$&$-3.12\times 10^{\text{-1}}$&$-2.604\times 10^{\text{-1}}$&$-2.312\times 10^{\text{-1}}$&$-2.12\times 10^{\text{-1}}$\\ \hline
$0.90$&$-7.863\times 10^{\text{-1}}$&$-3.432\times 10^{\text{-1}}$&$-2.588\times 10^{\text{-1}}$&$-2.199\times 10^{\text{-1}}$&$-1.975\times 10^{\text{-1}}$&$-1.808\times 10^{\text{-1}}$\\ \hline
$0.98$&$-5.003\times 10^{\text{-1}} $&$ - 2.753\times 10^{\text{-1}} $&$ -
              2.213\times 10^{\text{-1}} $&$ - 1.96\times 10^{\text{-1}} $&$ -
      1.805\times 10^{\text{-1}} $&$ - 1.699\times 10^{\text{-1}}$\\
 \hline
\end{tabular}

\vglue1.5cm
\begin{center}
{$\bf{ \sigma(u; \, w): ~ \alpha_S(m_Z) \, = \, 0.125}$ } \\*
\end{center}
\renewcommand{\tabcolsep}{0.26cm}
\begin{tabular}{|c||c|c|c|c|c|c|}
\hline \hline
$u$ &  $w=0.10$ & $w=0.28$ & $w=0.46$ & $w=0.64$ & $w=0.82$ & $w=1.00$ \\
\hline \hline
$0.01$&$2.309\times 10^{\text{-6}}$&$4.039\times 10^{\text{-6}}$&$5.47\times 10^{\text{-6}}$&$8.764\times 10^{\text{-6}}$&$2.845\times 10^{\text{-5}}$&$1.247\times 10^{\text{-4}}$\\ \hline
$0.02$&$2.46\times 10^{\text{-6}}$&$8.166\times 10^{\text{-6}}$&$1.706\times 10^{\text{-4}}$&$1.785\times 10^{\text{-3}}$&$9.505\times 10^{\text{-3}}$&$3.339\times 10^{\text{-2}}$\\ \hline
$0.03$&$2.641\times 10^{\text{-6}}$&$2.309\times 10^{\text{-4}}$&$5.49\times 10^{\text{-3}}$&$3.839\times 10^{\text{-2}}$&$1.458\times 10^{\text{-1}}$&$3.871\times 10^{\text{-1}}$\\ \hline
$0.04$&$5.702\times 10^{\text{-6}}$&$2.809\times 10^{\text{-3}}$&$4.39\times 10^{\text{-2}}$&$2.243\times 10^{\text{-1}}$&$6.666\times 10^{\text{-1}}$&$1.448\
$\\ \hline
$0.05$&$4.474\times 10^{\text{-5}}$&$1.562\times 10^{\text{-2}}$&$1.727\times 10^{\text{-1}}$&$6.899\times 10^{\text{-1}}$&$1.695\
$&$3.153$\\ \hline
$0.06$&$2.796\times 10^{\text{-4}}$&$5.422\times 10^{\text{-2}}$&$4.501\times 10^{\text{-1}}$&$1.47\
$&$3.092$&$5.072\
$\\ \hline
$0.07$&$1.174\times 10^{\text{-3}}$&$1.387\times 10^{\text{-1}}$&$9.02\times 10^{\text{-1}}$&$2.485\
$&$4.589$&$\
6.778$\\ \hline
$0.08$&$3.689\times 10^{\text{-3}}$&$2.875\times 10^{\text{-1}}$&$1.512\
$&$3.596$&$\
5.938$&$8.023\
$\\ \hline
$0.09$&$9.412\times 10^{\text{-3}}$&$5.117\times 10^{\text{-1}}$&$2.231\
$&$4.665$&$\
6.988$&$8.735\
$\\ \hline
$0.10$&$2.055\times 10^{\text{-2}}$&$8.124\times 10^{\text{-1}}$&$2.996\
$&$5.587$&$\
7.678$&$8.96\
$\\ \hline
$0.11$&$3.98\times 10^{\text{-2}}$&$1.181\
$&$3.745$&$\
6.303$&$8.017\
$&$8.791$\\ \hline
$0.12$&$7.011\times 10^{\text{-2}}$&$1.603\
$&$4.429$&$\
6.79$&$8.05\
$&$8.34$\\ \hline
$0.13$&$1.144\times 10^{\text{-1}}$&$2.057\
$&$5.012$&$\
7.056$&$7.838\
$&$7.71$\\ \hline
$0.14$&$1.752\times 10^{\text{-1}}$&$2.524\
$&$5.473$&$\
7.121$&$7.447\
$&$6.985$\\ \hline
$0.15$&$2.547\times 10^{\text{-1}}$&$2.982\
$&$5.805$&$\
7.019$&$6.935\
$&$6.228$\\ \hline
$0.16$&$3.542\times 10^{\text{-1}}$&$3.415\
$&$6.01$&$\
6.781$&$6.354$&$\
5.482$\\ \hline
$0.17$&$4.742\times 10^{\text{-1}}$&$3.809\
$&$6.098$&$\
6.443$&$5.743\
$&$4.777$\\ \hline
$0.18$&$6.145\times 10^{\text{-1}}$&$4.152\
$&$6.082$&$\
6.034$&$5.133\
$&$4.127$\\ \hline
$0.19$&$7.741\times 10^{\text{-1}}$&$4.439\
$&$5.975$&$\
5.582$&$4.544\
$&$3.54$\\ \hline
$0.20$&$9.512\times 10^{\text{-1}}$&$4.666\
$&$5.794$&$\
5.108$&$3.99\
$&$3.017$\\ \hline
$0.21$&$1.143\
$&$4.832$&$\
5.554$&$4.63\
$&$3.477$&$\
2.558$\\ \hline
$0.22$&$1.348\
$&$4.939$&$\
5.268$&$4.161\
$&$3.011$&$\
2.156$\\ \hline
$0.23$&$1.561\
$&$4.99$&$\
4.95$&$3.71$&$\
2.591$&$1.808\
$\\ \hline
$0.24$&$1.781\
$&$4.989$&$\
4.611$&$3.285\
$&$2.216$&$\
1.507$\\ \hline
$0.25$&$2.002\
$&$4.942$&$\
4.26$&$2.889\
$&$1.883$&$\
1.247$\\ \hline
$0.26$&$2.223\
$&$4.855$&$\
3.907$&$2.525\
$&$1.59$&$\
1.024$\\ \hline
$0.27$&$2.438\
$&$4.731$&$\
3.558$&$2.192\
$&$1.333$&$\
8.328\times 10^{\text{-1}}$\\ \hline
$0.28$&$2.647\
$&$4.577$&$\
3.217$&$1.891\
$&$1.108$&$\
6.686\times 10^{\text{-1}}$\\ \hline
$0.29$&$2.844\
$&$4.399$&$\
2.89$&$1.619\
$&$9.118\times 10^{\text{-1}}$&$5.278\times 10^{\text{-1}}$\\ \hline
$0.30$&$3.029\
$&$4.199$&$\
2.579$&$1.376\
$&$7.41\times 10^{\text{-1}}$&$4.067\times 10^{\text{-1}}$\\ \hline
$0.31$&$3.199\
$&$3.984$&$\
2.286$&$1.159\
$&$5.926\times 10^{\text{-1}}$&$3.022\times 10^{\text{-1}}$\\ \hline
$0.32$&$3.352\
$&$3.757$&$\
2.012$&$9.665\times 10^{\text{-1}}$&$4.639\times 10^{\text{-1}}$&$2.117\times 10^{\text{-1}}$\\ \hline
$0.33$&$3.486\
$&$3.522$&$\
1.758$&$7.956\times 10^{\text{-1}}$&$3.523\times 10^{\text{-1}}$&$1.334\times 10^{\text{-1}}$\\ \hline
$0.34$&$3.601\
$&$3.282$&$\
1.523$&$6.446\times 10^{\text{-1}}$&$2.557\times 10^{\text{-1}}$&$6.584\times 10^{\text{-2}}$\\ \hline
$0.35$&$3.696\
$&$3.04$&$\
1.308$&$5.114\times 10^{\text{-1}}$&$1.719\times 10^{\text{-1}}$&$7.763\times 10^{\text{-3}}$\\ \hline
$0.36$&$3.769\
$&$2.8$&$\
1.112$&$3.943\times 10^{\text{-1}}$&$9.927\times 10^{\text{-2}}$&$-4.2\times 10^{\text{-2}}$\\ \hline
$0.37$&$3.822\
$&$2.563$&$\
9.342\times 10^{\text{-1}}$&$2.914\times 10^{\text{-1}}$&$3.593\times 10^{\text{-2}}$&$-8.451\times 10^{\text{-2}}$\\ \hline
$0.38$&$3.855\
$&$2.331$&$\
7.732\times 10^{\text{-1}}$&$2.013\times 10^{\text{-1}}$&$-1.942\times 10^{\text{-2}}$&$-1.207\times 10^{\text{-1}}$\\ \hline
$0.39$&$3.866\
$&$2.106$&$\
6.284\times 10^{\text{-1}}$&$1.224\times 10^{\text{-1}}$&$-6.775\times 10^{\text{-2}}$&$-1.515\times 10^{\text{-1}}$\\ \hline
$0.40$&$3.858\
$&$1.89$&$\
4.984\times 10^{\text{-1}}$&$5.351\times 10^{\text{-2}}$&$-1.098\times 10^{\text{-1}}$&$-1.775\times 10^{\text{-1}}$\\ \hline
$0.50$&$2.917\
$&$2.99\times 10^{\text{-1}}$&$-2.15\times 10^{\text{-1}}$&$-2.945\times 10^{\text{-1}}$&$-3.031\times 10^{\text{-1}}$&$-2.883\times 10^{\text{-1}}$\\ \hline
$0.60$&$1.277\
$&$-3.637\times 10^{\text{-1}}$&$-3.894\times 10^{\text{-1}}$&$-3.592\times 10^{\text{-1}}$&$-3.194\times 10^{\text{-1}}$&$-2.902\times 10^{\text{-1}}$\\ \hline
$0.70$&$-6.465\times 10^{\text{-2}}$&$-5.265\times 10^{\text{-1}}$&$-4.05\times 10^{\text{-1}}$&$-3.365\times 10^{\text{-1}}$&$-2.936\times 10^{\text{-1}}$&$-2.656\times 10^{\text{-1}}$\\ \hline
$0.80$&$-7.947\times 10^{\text{-1}}$&$-4.951\times 10^{\text{-1}}$&$-3.589\times 10^{\text{-1}}$&$-2.94\times 10^{\text{-1}}$&$-2.578\times 10^{\text{-1}}$&$-2.344\times 10^{\text{-1}}$\\ \hline
$0.90$&$-9.333\times 10^{\text{-1}}$&$-3.996\times 10^{\text{-1}}$&$-2.936\times 10^{\text{-1}}$&$-2.452\times 10^{\text{-1}}$&$-2.179\times 10^{\text{-1}}$&$-1.978\times 10^{\text{-1}}$\\ \hline
$0.98$&$-5.886\times 10^{\text{-1}} $&$ - 3.088\times 10^{\text{-1}} $&$ -
              2.428\times 10^{\text{-1}} $&$ - 2.128\times 10^{\text{-1}}
$&$ -
      1.947\times 10^{\text{-1}} $&$ - 1.826\times 10^{\text{-1}}$\\
 \hline
\end{tabular}
\vglue .5cm



\end{document}